\documentclass[%
 reprint,
 amsmath,amssymb,
 aps,
]{revtex4-2}

\usepackage{graphicx}
\usepackage{dcolumn}
\usepackage{bm}

\usepackage{xcolor}
\usepackage{soul}

\newcommand{\f}{\frac}

\newcommand{\p}{\partial}
\newcommand{\de}{\delta}

\newcommand{\om}{\omega}

\newcommand{\ta}{\theta}

\newcommand{\la}{\lambda}
\newcommand{\La}{\Lambda}

\newcommand{\s}{\,\,\,}

\newcommand{\grad}{\vec{\nabla}}

\newcommand{\dotp}[2]{\vec{#1}\cdot\vec{#2}}

\begin{document}

\preprint{APS/123-QED}

\title{Field Theory of Birhythmicity}

\author{Sergei Shmakov}
 \email{sshmakov@uchicago.edu}
 \affiliation{James Franck Institute and Department of Physics, The University of Chicago, Chicago, Illinois 60637, United States
}
\author{Peter B. Littlewood}
 \email{littlewood@uchicago.edu}
\affiliation{James Franck Institute and Department of Physics, The University of Chicago, Chicago, Illinois 60637, United States
}
\affiliation{School of Physics and Astronomy, The University of St Andrews, St Andrews, KY16 9AJ, United Kingdom}

\date{\today}

\begin{abstract}
Non-equilibrium dynamics are present in many aspects of our lives, ranging from microscopic physical systems to the functioning of the brain. What characterizes stochastic models of non-equilibrium processes is the breaking of the fluctuation-dissipation relations as well as the existence of non-static stable states, or phases. A prototypical example is a dynamical phase characterized by a limit cycle - the order parameter of finite magnitude rotating or oscillating at a fixed frequency. Consequently, birhythmicity, where two stable limit cycles coexist, is a natural extension of the simpler single limit cycle phase. Both the abundance of real systems exhibiting such states as well as their relevance for building our understanding of non-equilibrium phases and phase transitions are strong motivations to build and study models of such behavior. Field theoretic tools can be used to provide insights into either phase and the transition between them. In this work we explore a simple linear model of the single limit cycle phase with phase-amplitude coupling. We demonstrate how such non-equilibrium coupling affects the fluctuation spectrum of the theory. We then extend this model to include a continuous transition to a two-cycle phase. We give various results, such as an appearance of a critical exceptional point, the destruction of the transition, enhancement of noise for the phase and the presence of KPZ dynamics. Finally, we qualitatively demonstrate these results with numerics and discuss future directions.

\end{abstract}

\maketitle

\section{\label{sec:intro}Introduction}

In equilibrium statistical physics the study of phases and phase transitions has been a long-standing field that provided innumerable insights. Among those, the ideas of universality and renormalization group have been profoundly powerful and useful \cite{Altland2023, Tauber2014, Kamenev2023}. They allow one to classify the phases of matter into general categories, universality classes, that allow understanding of real and complicated systems from studying much simpler models belonging to the same class. Non-equilibrium conditions open the space to a wide variety of new phases and classes. There are tremendous efforts to build tools, field theories in particular, akin to the equilibrium ones to understand these phases and the respective transitions \cite{Altland2023,Tauber2014,Kamenev2023,Cross1993, Zelle2023, Daviet2024}. Similarly to equilibrium studies, it should prove valuable to develop the simplest models possible that can describe those. From the perspective of dynamical systems one of the rudimentary steady states that are not purely relaxational are periodic orbits, or  limit cycles. They can therefore be taken as paradigmatic states that one can use to build understanding of non-equilibrium phases of matter. For instance, a transition from a both an equilibrium disordered and equilibrium ordered states of $O(N)$ models to limit cycle phases is already being extensively studied \cite{Zelle2023, Daviet2024}.

Periodic orbits, oscillations and limit cycles are non-equilibrium features of a broad range of systems appearing in many fields, varying from the studies of population dynamics, biological sciences and neuroscience, and physics \cite{Strogatz2024, Guckenheimer2013, Kuramoto2003, Pikovskij2007, Tian2022}. Their presence in all these fields makes their study important in its own right, in addition to the more theoretical reasons outlined above. Birhythmicity is a phenomenon where the system supports two stable frequencies, or limit cycles. Although the parameter space where it happens can be small, birhythmicity occurs in various fields, including physics \cite{Goldstone1984, Ritter1993, Biswas2016}, biology and biochemistry \cite{Goldbeter2018, Decroly1982, Tsumoto2006, Fuente1999}, control theory and mathematical modeling \cite{Biswas2016, Yamapi2010, Kar2004}, among many others. In many of these examples, in fact, there is a transition from a single limit cycle to a birhythmic phase. 

In particular, oscillations are a fundamental part of brain dynamics, both healthy and afflicted by disease \cite{Tian2022, Cook2022, Basar2013, diSanto2018, Schroeder2009, Slepukhina2023, Wilson1972}. As an example, dynamic modeling of a transition to an epileptic seizure often utilizes a transition to a stable limit cycle \cite{Milton2010, Junges2020, Harrington2024}. From a different perspective a synchronization transition between oscillating units has been proposed as an explanation of brain criticality \cite{diSanto2018}. It is therefore useful for the study of neural dynamics to construct simple, coarse-grained, albeit abstract, models in an attempt to understand general features underlying limit cycle phases and transitions between them. Thus, given the prevalence of oscillations and birhythmicity in real systems and the utility of limit cycles in developing models of non-equilibrium phase transitions, the focus of our work is to construct a generic field-theoretic model of a transition from mono- to birhythmicity.

Virtually any real system will be spatially extended and noisy. Thus, just like in classical studies of phase transitions, we need to include dimentionality (spatial extent) and fluctuations in order to fully characterize such a transition. Both of these ingredients can have profound effects on the behavior of limit cycles in particular. For example, spatially extended or lattice models of oscillators exhibit synchronization \cite{Acebrón2005, Rosenblum2004}, and noise plays important role in control theory \cite{Rosenblum2004, Tukhlina2008, Goldobin2003}, with many other important effects and applications not mentioned here. Dimensionality, of course, also plays a central role in understanding the universality class of a phase transition, and changes the associated critical exponents. In order to include these ingredients, as well as for preserving the most generality, we resort to using a generalized complex Landau-Ginzburg equation. Equations of this sort have a wide variety of applications and motivations, see e.g. \cite{Aranson2002, Cross1993, Hohenberg1977} and references therein. We add Gaussian white noise for simplicity.   

In noisy non-equilibrium systems, the fluctuation-dissipation relations don't hold \cite{Surowka2018, Sieberer2015, Kamenev2023, Tauber2014}. The phases and phase transitions occurring in such systems can belong to universality classes that don't have an equilibrium counterpart or don't belong to Hohenberg-Halperin classification \cite{Hohenberg1977, Tauber2014}. A paradigmatic example of such a model is the Kardar-Parisi-Zhang (KPZ) universality class \cite{Kardar1986, Kamenev2023}. Also relevant for the study of phase transitions is the symmetry group of the associated order parameter. A generic group that can be picked is $O(N)$, or $SO(N)$. Possible non-equilibrium phases of $O(N)$ models, as well as transitions between them, are being extensively studied in contexts similar to this letter \cite{Zelle2023, Daviet2024}, as mentioned above. For our model, we restrict to a two-component, or complex, field, meaning that we stick to $O(2)$ symmetry. 

Another aspect of such models, which is relevant for our system as well, is the possibility of exceptional points in parameter space. From a dynamical system's perspective, such points correspond to coalescence of dynamical modes with an associated merging of eigenvalues \cite{Weis2023, Fruchart2021}. These points can only occur outside of equilibrium \cite{Zelle2023}, and enhanced fluctuations or superthermal mode occupation are associated with exceptional points that also happen to be critical (CEP) \cite{Zelle2023, Hanai2020}. These points imply profound effects for the underlying system, and are also a focus of extensive studies \cite{Zelle2023,Hanai2020,Weis2023, Hanai2019,Chen2017, Zhang2019, Shmakov2024, Shmakov2024b}. 

The initial "toy" model describes a system that has a stable limit cycle in one phase which splits continuously into two stable limit cycles of different radii. The non-equilibrium feature is the coupling between the frequency of the cycles and their radii, which should be included for generality, but is a feature that often occurs in real systems \cite{Henry1982, Lauter2015, Canolty2010}. We show that this coupling leads directly to the transition happening through a CEP, at least at the mean field level. At the Gaussian level it leads to anomalously large contributions to the correlation function, which diverge as $k^{-6}$ in momentum space. That is contrasted with the usual feature of exceptional points where the correlations in the fluctuation spectrum diverge as $k^{-4}$ \cite{Zelle2023, Hanai2020}. Such a divergence has an effect on the models critical dimensions, which in turn dictate if real systems described by such models truly undergo phase transitions. It can also be understood as the origin of $1/f^n$ type noises in real systems, where $n$ depends on dimension. 

However, when the equation of motion for the field is written in terms of amplitude and phase, the simplest diffusive coupling generates terms that break the "$\mathcal{Z}_2$" symmetry around the finite radius where the splitting occurs. Although a symmetry breaking field can be introduced by hand, we argue that these terms are an inevitable contribution that destroy the continuous transition. In the language of dynamical systems, the transition instead occurs through a saddle-node bifurcation of limit cycles, which is a route that systems often take to reach birhythmicity \cite{Yamapi2010, Decroly1982, Biswas2016, Goldstone1984}. However, unlike the usual external field, these terms can have unexpected consequences from the renormalization group perspective. We only briefly mention these ideas here, and reserve more detailed analysis for future work. 

This paper is organized as follows. We present a general Landau-Ginzburg model in section \ref{sec:model}. Section \ref{sec:linear} discusses the linearized theory about mean-field (the equivalent to the gaussion theory in equilibrium phase transitions). Section \ref{sec:continuous} explores the simplest model for the transition from a single stable orbit to a pair of limit cycles, and is generalized in section \ref{sec:full} to a nonlinear scaling theory, that demonstrates the general breaking of $Z_2$ symmetry. This implies the generation of 1/f like noise down to some cutoff. Numerical simulations in section \ref{sec:sims} support the earlier scaling analysis and demonstrate droplet growth (in 2D) characteristic of a weakly first-order-like transition.

\section{\label{sec:model}General model}
We begin with a general Landau-Ginzburg equation for a complex field $\psi$ that would include all of the ingredients discussed in the Introduction. Such a model could be written as:
\begin{equation}
    \p_t \psi = f_1(|\psi|)\psi + i f_2(|\psi|)\psi + D\nabla^2 \psi + \xi \,.
\end{equation}

This equation has an $O(2)$ symmetry with respect to the field $\psi = \psi_x + i\psi_y$. Here we will only focus on rotations, which is the same as considering the $U(1)$ symmetry only. The real function $f_1(|\psi|)$ is responsible for producing multistability in the radial direction, and only depends on the amplitude of the field in order to preserve the symmetry. The real function $f_2(|\psi|)$ is the coupling between the frequency of the rotation of the field and it's amplitude. In the language of Langevin dynamics, $f_2$ is responsible for breaking the symmetry conditions that define equilibrium. This can be shown by breaking up this equation into components and directly checking for the symmetry defining thermodynamic equilibrium \cite{Sieberer2015, Zelle2023}, or by using the ghost fields and supersymmetry with the associated BRST-like transformations \cite{Hertz2016, Surowka2018}. Two component order parameters with $O(2)$ symmetry offer a unique simplicity to study non-equilibrium systems. For a general $O(N)$, in order to go out of equilibrium, one needs to supply many additional terms that render the dynamics more complex than overdamped relaxation \cite{Zelle2023}. For $O(2)$, however, any rotation commutes with any other rotation. Thus, complex terms (such as $if_2 \psi$) which break equilibrium conditions are easy to generate while preserving the overall symmetry.

We implement spatial extent in the form of a diffusion term with strength $D$. This is conceptually the simplest way to do so, but diffusion is of course present in all sorts of continuous descriptions of real systems \cite{Cross1993, Hohenberg1977, Aranson2002}. Although generically $D$ can be taken to have both real and imaginary parts (e.g. for models of condensate dynamics these would correspond to coherent and dissipative couplings), we take it to be real for simplicity. Finally, we add white Gaussian noise $\xi$ with correlations $\left<\xi(t,\vec{x})\xi(t',\vec{x}')\right> = 0 = \left<\xi^*(t,\vec{x})\xi^*(t',\vec{x}')\right>$ and $\left<\xi(t,\vec{x})\xi^*(t',\vec{x}')\right> = 2B \de(t-t')\de^d(\vec{x}-\vec{x}')$, where $B$ is the noise strength. Such a model is an example of a $\la$-$\om$ system in the language of \cite{Aranson2002, Cross1993}, and has a wide range of applications (see references therein).

Defining the amplitude-phase representation through $\psi(t,\vec{x}) = r(t,\vec{x})e^{i\ta(t,\vec{x})}$, after carefully keeping track of the subtleties due to Ito discretization \cite{Kamenev2023, Gardiner1997, Hertz2016}, the original equation can be written as:
\begin{equation}
    \begin{split}
        \p_t r &= f_1(r)r + D\nabla^2 r - D r\grad \ta \cdot \grad \ta + \f{B}{2r} + \xi_r\\
        \p_t \ta & = f_2(r) + D\nabla^2 \ta +\f{2D}{r} \grad \ta \cdot \grad r + \f{1}{r}\xi_\ta
    \end{split} \,.
\end{equation}

Here the noise terms are decorrelated and each have strength $B$: $\left<\xi_i(t,\vec{x})\xi_j(t',\vec{x}')\right> = B \de_{ij}\de(t-t')\de^d(\vec{x}-\vec{x}')$. We also pick up a centrifugal term, as well as two terms generated by diffusion. The former can be absorbed into $f_1(r)$ and is less relevant for our considerations below, while the latter will be addressed in the following sections. This is as far as general considerations take us. 

\section{\label{sec:linear}Linear theory}
Suppose at this point that we are in a regime with a single limit cycle state. In that case the function $f_1(r)$ produces a single stable point for the dynamics of $r$. As the first intuitive step in the study of dynamical systems and field theories, we linearize around that point. Calling the stable radius $r_{st} = a$, we define $r = a + \de r$. Assuming the stability is controlled by parameter $m$ (referencing mass or reduced temperature terms in field theories), we get:
\begin{equation}
    \p_t \de r = -m \de r + D \nabla^2 \de r + \xi_r \,.
\end{equation}

We then keep linear terms for the evolution of the phase, assuming $f_2(r) \approx f_2(r_{st}) + F \de r$, where we defined $F = \f{d f_2}{d r}(r_{st})$, and we also ignore $f_2(r_{st})$ by going into a rotating frame:
\begin{equation}
    \p_t \ta = F \de r + D\nabla^2 \ta + \f{1}{a}\xi_\ta \;.
\end{equation}

To obtain the fluctuation spectrum of such a system, we employ the Martin-Siggia-Rose-Janssen-De Dominicis procedure \cite{Tauber2014,Altland2023,Kamenev2023, Martin1973, DeDominicis1978} with response fields $\bar{r}$, $\bar{\ta}$. It will prove useful further down in our considerations. The dynamical action produced by these equations is:
\begin{equation}
    \begin{split}
        &S_0 = \int dt d^dx \, i\bar{r}\left(\p_t \de r+ m \de r - D_r\nabla^2 \de r \right) \\&+i\bar{\ta}\left(\p_t \ta - F\de r - D_\ta \nabla^2 \ta\right) - \f{B_r}{2}\bar{r}^2 -\f{B_\ta}{2}\bar{\ta}^2 \; ,
    \end{split} 
\end{equation}

where we defined $B_r = B$, $B_\ta = B/a^2$ and $D_r = D_\ta = D$ in anticipation that the renormalization group flow will change these couplings from these initial conditions. The correlators in Fourier space (which we define as $v(t,\vec{x}) = \int \f{d\om d^dk}{(2\pi)^{d+1}} \hat{v}(\om,\vec{k})e^{i\om t - i\dotp{k}{x}}$) are given by $G_0(\om,\vec{k} ;\om',\vec{k}' ) = (2\pi)^{d+1}G_0(\om,\vec{k})\de(\om+\om')\de^d(\vec{k}+\vec{k}')$, with:
\small\begin{widetext}
\begin{equation}
    \begin{split}
        G_0(\om,\vec{k}) &= \begin{pmatrix}
        \left<\hat{r}(\om,\vec{k})\hat{r}(-\om,-\vec{k})\right>_0 & \left<\hat{r}(\om,\vec{k})\hat{\bar{r}}(-\om,-\vec{k})\right>_0 & \left<\hat{r}(\om,\vec{k})\hat{\ta}(-\om,-\vec{k})\right>_0 & 0 \\
         \left<\hat{\bar{r}}(\om,\vec{k})\hat{r}(-\om,-\vec{k})\right>_0 & 0 & \left<\hat{\bar{r}}(\om,\vec{k})\hat{\ta}(-\om,-\vec{k})\right>_0 & 0 \\
         \left<\hat{\ta}(\om,\vec{k})\hat{r}(-\om,-\vec{k})\right>_0 & \left<\hat{\ta}(\om,\vec{k})\hat{\bar{r}}(-\om,-\vec{k})\right>_0 & \left<\hat{\ta}(\om,\vec{k})\hat{\ta}(-\om,-\vec{k})\right>_0 & \left<\hat{\ta}(\om,\vec{k})\hat{\bar{\ta}}(-\om,-\vec{k})\right>_0 \\ 0& 0& \left<\hat{\bar{\ta}}(\om,\vec{k})\hat{\ta}(-\om,-\vec{k})\right>_0 &0
    \end{pmatrix} \\&= \begin{pmatrix}
        \f{B_r}{\om^2 + (m+Dk^2)^2} & \f{1}{\om-im-iD_rk^2} & \f{iB_rF}{(\om+iD_\ta k^2)(\om^2+(m+D_rk^2)^2)} & 0 \\[12pt]
        \f{1}{-\om-im-iD_rk^2} & 0 &\f{iF}{(\om+iD_\ta k^2)(-\om-im-iD_rk^2)} & 0 \\[12pt]
        \f{iB_rF}{(-\om+iD_\ta k^2)(\om^2+(m+D_rk^2)^2)} &  \f{iF}{(-\om+iD_\ta k^2)(\om-im -iD_r k^2)} & \f{B_\ta}{\om^2+D_\ta ^2 k^4}+\f{B_r F^2}{(\om^2+D_\ta ^2 k^4)(\om^2+(m+D_rk^2)^2)} & \f{1}{\om - i D_\ta k^2} \\[12pt]
        0&0&\f{1}{-\om-iD_\ta k^2} & 0 
        \end{pmatrix} \, .
    \end{split}
\end{equation}    
\end{widetext}
\normalsize
As can be seen the coupling $F$ introduces cross-correlations between the radial and the phase degree of freedom. One way to detect a continuous phase transition is to look at the equal time correlation function. In this case, the Keldysh components of the correlator \cite{Kamenev2023,Altland2023,Tauber2014}, which do not involve the response fields, are given by:
\begin{equation}
\begin{split}
    &G^K_{0,rr}(t=0,\vec{k}) = \int\f{d\om}{2\pi} \left<\hat{r}(\om,\vec{k})\hat{r}(-\om,-\vec{k})\right>_0 \\&=\f{B_r/2}{m+D_r k^2} \; , \\
    &G^K_{0,r\ta}(t=0,\vec{k}) = \int\f{d\om}{2\pi} \left<\hat{r}(\om,\vec{k})\hat{\ta}(-\om,-\vec{k})\right>_0\\&=\f{B_rF/2}{(m+D_r k^2)\left[m+(D_r+D_\ta)k^2\right]} \; , \\
     &G^K_{0,\ta\ta}(t=0,\vec{k}) = \int\f{d\om}{2\pi} \left<\hat{\ta}(\om,\vec{k})\hat{\ta}(-\om,-\vec{k})\right>_0\\&=\f{B_\ta/2}{D_\ta k^2}+\f{B_rF^2/2}{D_\ta k^2(m+D_r k^2)\left[m+(D_r+D_\ta)k^2\right]} \; . \\
\end{split} 
\end{equation}

The radial degree of freedom has a correlation function that is associated with the Ising or Model A second order transition \cite{Tauber2014,Kamenev2023,Hohenberg1977}. The phase correlation function has a $k^{-2}$ divergence at low momenta, also typical for Goldstone modes in equilibrium \cite{Tauber2014,Kamenev2023,Hohenberg1977, Altland2023}. Even though we have a non-equilibrium coupling $F$ that changes the precise form of correlation function, we will only see deviations from the usual Goldstone mode behavior when the radial degree of freedom becomes critical at $m=0$, or for high momenta $k^2>> m/D_r$. The off-diagonal term is also damped unless we are at the critical point. We will discuss this point below.

At high momenta, when $k^2 >> m/D_r$, the correlation function for the phase degree of freedom behaves as $k^{-6}$. Thus, there is a length scale $l_c \sim \sqrt{D_r/m}$ below which the phase correlation function for a system of size $L>>l_c$ behaves as:
\begin{equation}
    \left<e^{i\ta(0,\vec{x}_1)-i\ta(0,\vec{x}_2)}\right> \sim \exp \left[-\f{C_{d,x} B_r F^2 L^{4-d}}{D_r D_\ta (D_r+D_\ta)}|\vec{x}_1-\vec{x}_2|^2\right] \,.
\end{equation}

Similarly, there is a critical time scale $t_c \sim m^{-1}$, below which the time correlation function picks up a new dependence on time:
\begin{equation}
    \left<e^{i\ta(t_1,0)-i\ta(t_2,0)}\right> \sim \begin{cases}
        &\exp \left[-C_{d,t} |t_1-t_2|^2\right] \;, d=1,2\\
        &\exp \left[-C_{d,t} |t_1-t_2|^{\f{3}{2}}\right] \;, d=3
    \end{cases} \;.
\end{equation}

Where $C_{d, x/t}$ are some constants with appropriate combinations of $F$, $B_r$ $D_r$, $D_\ta$ and system size $L$. Both these scales can be understood by considering the $F\de r$ term as an additional Ornstein-Uhlenbeck, spatially correlated noise \cite{Altland2023}. In that case we see that it is simply another source of fluctuations that destroys phase order on the scales defined by it's time and spatial correlation lengths. We demonstrate some of these effects in Sec. \ref{sec:sims}.

Another way to see this enhancement of fluctuations is to integrate out the radial degree of freedom. In that case, the phase noise is enhanced by the correlation function of the radial degree of freedom:
\begin{equation}
    B_\ta \rightarrow B_\ta + \f{B_r F^2}{\om^2+(m+D_r k^2)^2}
\end{equation}

This is reminiscent of the amplification of noise observed in \cite{Biancalani2017}. There the amplification also relied on the non-orthogonality of the eigenvectors of the linear connectivity matrix, which in the current linear case is given by (ignoring diffusive contributions):
\begin{equation}
        \p_t \begin{pmatrix}
            \de r \\ \ta
        \end{pmatrix} = - M \begin{pmatrix}
            \de r \\ \ta
        \end{pmatrix} = - \begin{pmatrix}
            m & 0 \\ -F & 0
        \end{pmatrix}\begin{pmatrix}
            \de r \\ \ta
        \end{pmatrix} \,.
\end{equation}

Another consequence of this amplification can be observed by integrating over the momenta in order to get a contribution local in space. In low, physical dimensions $d \leq 3$ and for $\om >> m$, this contribution becomes (aside from numerical constants):
\begin{equation}
    \int d^dk  \f{B_r F^2}{\om^2+(m+D_r k^2)^2} \sim \f{B_r F^2}{D_r^{\f{d}{2}}\om^{\f{4-d}{2}}} \;.
\end{equation}

Thus non-reciprocal coupling between degrees of freedom can be a source of colored, e.g. $1/f$, noise. The structure of correlators in non-reciprocally coupled linear systems is explored in \cite{Shmakov2024b}.

\section{\label{sec:continuous}Continuous transition}
\begin{figure}[!t]
\includegraphics[scale=0.38]{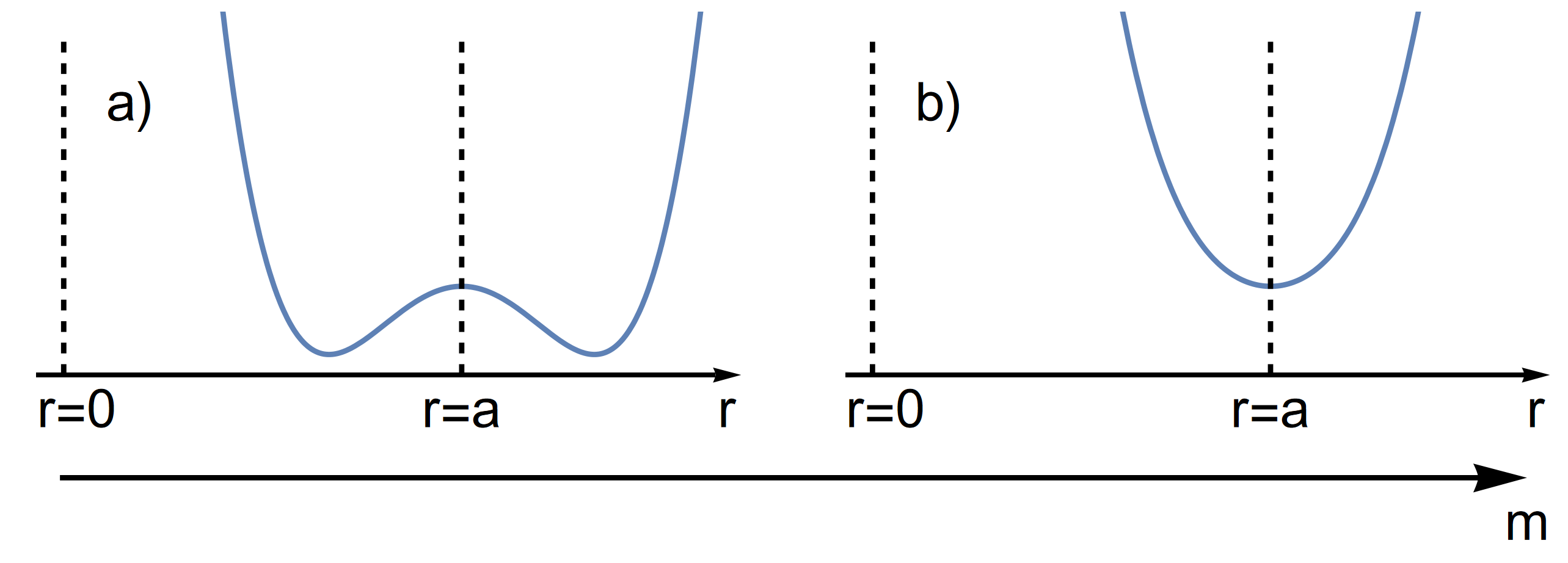}
\caption{\label{fig:TransitionSchematic} Schematic of the continuous transition model at mean field level. The curves indicated the "potential" along which the radial degree of freedom evolves towards the minima. a) Regime with two stable radii at which the limit cycles evolve ($m<0$ here at the mean field level). b) Regime with one stable radius for the limit cycle at $r=a$. For large enough $m$, this can be approximated by the linear theory. }
\end{figure}

With the linear theory understood, we can now explore transitions to birhythmicity. Although we saw above that there are non-linear terms in the phase-amplitude representation that produce more complex dynamics, we still believe it useful to understand a simpler toy model. We will assume that the two stable limit cycles appear continuously out of the original cycle through a pitchfork bifurcation, see Fig. \ref{fig:TransitionSchematic}. This is equivalent to the dynamical Ising or model A theory \cite{Tauber2014, Kamenev2023, Altland2023, Hohenberg1977} for the deviations of the radius $\de r$:
\begin{equation}
    \p_t \de r = -m \de r - g \de r^3 + D_r\nabla^2 \de r + \xi_r \;,
\end{equation}

where $m$ is the control parameter taking the system through a transition, and $g$ is some positive constant. Here it is useful consider this from the mean field perspective, and again look back at the linear theory. The eigenvalues and eigenvectors of matrix $M$ when $F\neq 0$ defined above are given by:
\begin{equation}
    \begin{split}
        &v_1 = \begin{pmatrix}
            0 \\ 1
        \end{pmatrix} \;,\; \la_1 =0 \\
        &v_2 = \begin{pmatrix}
            -\f{m}{F} \\ 1
        \end{pmatrix} \;,\; \la_2 =m
    \end{split} \;.
\end{equation}

We see that as we approach $m=0$, the eigenvalues become equal and eigenvectors coalesce. It can be checked that the same is true approaching the critical point from the other side with respect to each stable point. At the critical point the matrix is non-diagonalizable, making it a critical exceptional point (CEP). Exceptional points play an important role in dynamical systems studies \cite{Weis2023, Shmakov2024}, as well as in many-body and quantum physics \cite{Hanai2019,Hanai2020,Daviet2024, Zelle2023, Chen2017_2, Zhang2019_2, huang25}. 

One of the signatures of a CEP is a stronger divergence of fluctuations with momenta. A common enhancement usually reveals itself in a $k^4$ divergence of correlators \cite{Hanai2019,Hanai2020,Zelle2023}. We do see such fluctuations for the off-diagonal term $G^K_{0,r\ta} \sim k^{-4}$, where they often present. However, the phase fluctuation term diverges as $k^6$, which has not been observed before, to authors knowledge. The mechanism of this divergence lies in the phase being a Goldstone mode, which makes it "critical" a priori. Thus such a divergence is a result of tuning two fields to be critical and coupling them non-reciprocally. In this case, we see that the phase correlations decay as in eq. 8 at all distances, which destroys the usual behavior of an XY model. Similar to the Mermin-Wagner theorem \cite{Mermin1966}, this enhanced divergence implies that the ordering of the field will be destroyed in dimensions below $d=6$ at the CEP. 

It can be shown that a phase with dynamics defined by eq. 4 cannot exhibit a synchronization transition in low dimensions. This can be seen by defining a suitable Kuramoto order parameter \cite{Kuramoto2003}, or by observing the angle and phase correlation functions. However, at this level of description, the frequencies of oscillations do order. This can be seen, for example, by observing that $\left<\p_t \ta\right> = F\left<\de r\right>$, so when the radial degree of freedom orders, so does the frequency. This is also shown in Sec. \ref{sec:sims}.

\begin{figure}[!t]
\includegraphics[scale=0.23]{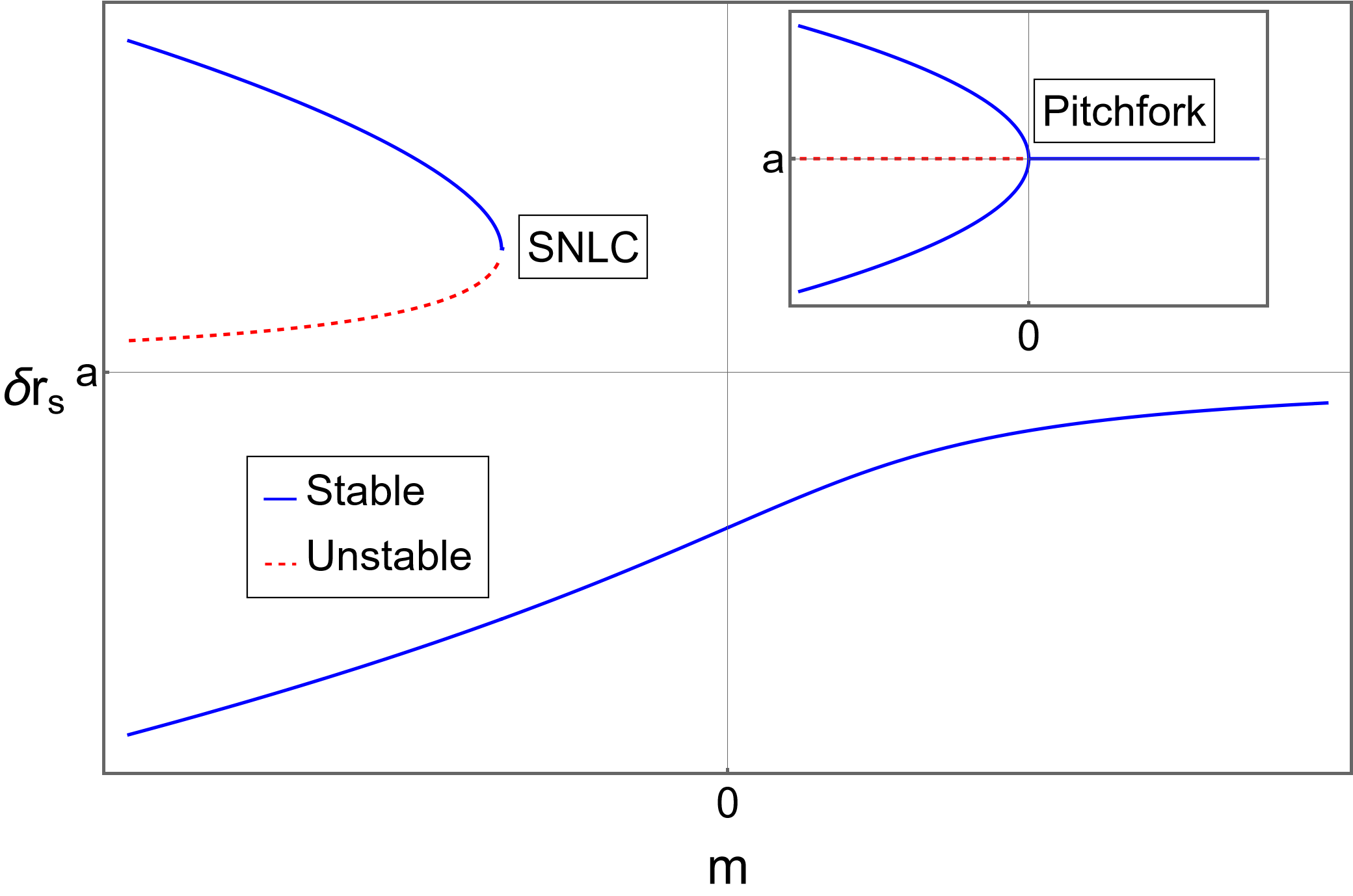}
\caption{\label{fig:Bifurcation} Saddle-node bifurcation scenario for the transition to birhythmicity. Solid blue lines correspond to stable radii, dashed red line corresponds to unstable radii. Either from the addition of an external field, or through the other nonlinearities described in the text, the transition happens through a saddle-node bifurcation of limit cycles (SNLC). The second stable radius (and therefore frequency) appears together with an unstable one as the control parameter $m$ is varied. The inset shows the continuous transition scenario described in sec. \ref{sec:continuous}, for comparison.}
\end{figure}

As mentioned before, however, often the route to birhythmicity is not continuous, but rather a saddle-node bifurcation \cite{Goldstone1984, Ritter1993, Biswas2016, Goldbeter2018, Decroly1982, Tsumoto2006, Yamapi2010}. If the mean field radial equation of motion is written as $\p_t \de r = -\de V(\de r)/ \de (\de r)$ for some potential $V$, the saddle node bifurcation corresponds to the appearance of a meta-stable well and a "hill" as the control parameter is tuned, see Fig. \ref{fig:Bifurcation}. In our case it can be modeled by an addition of an external field $h$ to the r.h.s. of eq. 13. Just like in usual statistical field theory, we then don't have a continuous transition between two phases, but rather a cross-over. In other words, there is no symmetry-broken phase. The problem can then be analyzed by usual means, i.e. through linearizing around the minima, using Kramer's estimation of the escape time \cite{Kamenev2023}, e.t.c. An analysis of a similar problem in zero dimensions was performed in \cite{Shmakov2024}. However, as mentioned above, the expansion of the full equation of motion inevitably leads to terms that break the $\mathcal{Z}_2$ symmetry for $\de r$ needed for a continuous transition.

We also argue that even if those terms are ignored, that symmetry will always be broken. The radial degree of freedom is defined in the domain $[0,\infty)$, or $[-a, \infty)$ for $\de r$. That means that it is impossible to construct a function $f_1(r) = f_1(a+\de r)$ that would not change under $\mathcal{Z}_2$ symmetry about a finite radius $a$, simply because it's not defined for $r<0$. The only option is to bound it from above as well, such that $f_1(r)$ is infinite at $r = 2a$.

\section{\label{sec:full} Full theory}
If we now consider all of the terms in eq. 2, we see that even at lowest order level in the expansion around the stable radius $a$, we get a highly non-linear action:
\begin{equation}
    \begin{split}
        &S = S_0 + S_{int} \;, \\
        &S_0 = \int dt d^dx \, i\bar{r}\left(\p_t \de r+ m \de r - D_r\nabla^2 \de r \right) \\&+i\bar{\ta}\left(\p_t \ta - F\de r - D_\ta \nabla^2 \ta\right) - \f{B_r}{2}\bar{r}^2 -\f{B_\ta}{2}\bar{\ta}^2 \; , \\
        &S_{int} = \int dt d^dx \, i\bar{r}\left(g\de r^3 + u \grad \ta \cdot \grad \ta \right) +iv\bar{\ta} \grad \ta \cdot \grad \de r \; . 
    \end{split} 
\end{equation}

Here we defined $u = Da$ and $v = 2D/a$, again in the anticipation of the RG flow taking those couplings away from their initial values. The usual $\mathcal{Z}_2$ symmetry of the radial degree of freedom $(\de r, \bar{r}) \rightarrow -(\de r, \bar{r})$ is broken by the $iu\bar{r}\grad \ta \cdot \grad \ta$ term. For the sake of the radial transition, it acts as an external field, which can be seen by the lowest order replacement:
\begin{equation}
    \int dt d^dx\, iu\bar{r} \grad{\ta}\cdot\grad\ta \rightarrow \int dt d^dx\, iu\bar{r} \left<\grad{\ta}\cdot\grad\ta\right> \; .
\end{equation}

However, this is a crude approximation, and a better way to asses the contribution of this term is through an RG procedure. Although we do not attempt to perform the complete procedure here, we can gain insight from the "zero-loop" or scaling arguments that are part of RG. Following standrad dynamic renormalization group procedures \cite{Kamenev2023, Tauber2014, Altland2023}, we want to only keep the slow momenta $|k|<\La/b$, where $b>1$, and then scale the coordinates and the fields as:
\begin{equation}
    \begin{split}
        &\vec{x} \rightarrow b\vec{x} \s,\s\s t \rightarrow b^zt \\
    & r (t, \vec{x}) \rightarrow b^{\chi_r} r (b^zt, b\vec{x}) \s,\s\s \bar{r} (t, \vec{x}) \rightarrow b^{\bar{\chi}_r}\bar{r} (b^zt, b\vec{x}) \\
    &\ta (t, \vec{x}) \rightarrow \ta(b^zt, b\vec{x}) \s,\s\s  \bar{\ta} (t, \vec{x}) \rightarrow b^{\bar{\chi}_\ta}\bar{\ta} (b^zt, b\vec{x}) \;.
    \end{split}
\end{equation}

Notice that the phase field is not scaled in order to preserve it's compactness and the orientation of the order parameter. We then pick the scalings of the fields as to preserve the free part of the Gaussian action for $r$ and the diffusion term for $\ta$, which yields $z=2$, $\chi_r = 1-\f{d}{2}$, $\bar{\chi}_r = -1-\f{d}{2}$, $\bar{\chi}_\ta = -d$. The scaling of the radial fields match the usual second order critical point scaling \cite{Kamenev2023}. With this, the couplings that are changed due to scaling are:
\begin{equation}
    \begin{split}
        &m \rightarrow b^2 m \\
        &F \rightarrow b^{3-\f{d}{2}}F\\
        &B_\ta \rightarrow b^{2-d}B_\ta
    \end{split}
    \quad ,\quad
    \begin{split}
        &g \rightarrow b^{4-d}g\\
        & u \rightarrow b^{\f{d}{2}-1}u \\
        & v \rightarrow b^{1- \f{d}{2}}v \\
    \end{split} \;.
\end{equation}

Although we have a symmetry breaking term, at least at the level of scaling it is irrelevant in $d=1$. Similarly, the other term $iv\bar{\ta}\grad \ta \cdot \grad \de r$ is relevant in low dimensions, although it is symmetric under $\mathcal{Z}_2$. Moreover, if they are considered without any approximations, as in eq. 2), both of these terms are marginal. All this brings a possibility of a symmetry restoration for the radial field at long distances. And, of course, the term $F$ that pushes the system out of equilibrium is highly relevant for $d<6$. We also see that the phase noise can be irrelevant in $d>2$, which means that the phase will only fluctuate through it's coupling to the radius, and thus might experience more ordering. Seeing also that the bare value of the phase noise is given by $B_r/a^2$, it might be appropriate to set the phase noise to zero in high dimensions. However, a more detailed analysis is needed to assert what actually occurs in each dimension.

Additionally, new terms will be generated by the RG flow. For instance, since we can have propagators between $\ta$ and $\de r$, we can generate a KPZ contribution to the dynamics of the angle. That can be seen by combining the $u$ and $F$ vertices. Even outside of RG, if we take the part of the radial action linear in $\de r$ and integrate it out, we get a contribution to the phase degree of freedom that is proportional to:
\begin{equation}
\begin{split}
    \sim& -Fu\int_{x,x',t,t'} \bar{\ta}(x,t)(\grad \ta)^2(x',t')\\&\left[G^A_{0,rr}(x-x',t-t') + G^R_{0,rr}(x'-x,t'-t)\right]   
\end{split} \;.
\end{equation}

Here $G^{A/R}_{0,rr}$ are the advanced and retarded propagators of the radial degree of freedom, equal to $\left<r \bar{r}\right>$. Although in equilibrium the local part of that expression is zero \cite{Kamenev2023}, it might not be so outside of equilibrium, where the usual identities between correlators might not hold. Either way, we see that the existence of this term will be based on non-zero $F$, which is expected since KPZ dynamics is non-equilibrium. In fact, since we don't have the additional $\ta\rightarrow -\ta$ (with the associated $\bar{\ta}\rightarrow -\bar{\ta}$) symmetry due to the presence of the non-equilibrium term, KPZ scaling is expected for limit cycles phases \cite{Daviet2024b}.

In the presence of $F$, other generated terms will include a $\de r^2$ non-linearity, cross-diffusions between the radial and angular degree of freedom, cross-correlated noise, etc. All of these can be observed even at a one-loop level.

\section{\label{sec:sims}Simulations}
To investigate the theoretical predictions outlined above we perform lattice simulations of the respective stochastic processes. In order to take into account the compactness of the phase field, we replace the spatial derivatives as:
\begin{equation}
\begin{split}
    \nabla^2 \ta &\approx \sum_{\hat{a}} \sin \left[\ta(\vec{x}+\hat{a}) - \ta(\vec{x})\right] \\
    \left(\grad \ta \right)^2 &\approx \sum_{\hat{a}}\left(1-\cos \left[\ta(\vec{x}+\hat{a}) - \ta(\vec{x})\right]\right)
\end{split} \;,
\end{equation}

where $\vec{x}$ are the locations of the lattice points and the sum  $\sum_{\hat{a}}$ goes over the nearest neighbors of $\vec{x}$, where $\hat{a}$ are lattice vectors. This is in accordance to studies of compact KPZ equation \cite{Sieberer2016}, here extended to arbitrary dimensions. To demonstrate the theoretical results we only keep the $u\grad \ta \cdot \grad \ta$ term. 

\begin{figure}[!t]
\includegraphics[scale=0.22]{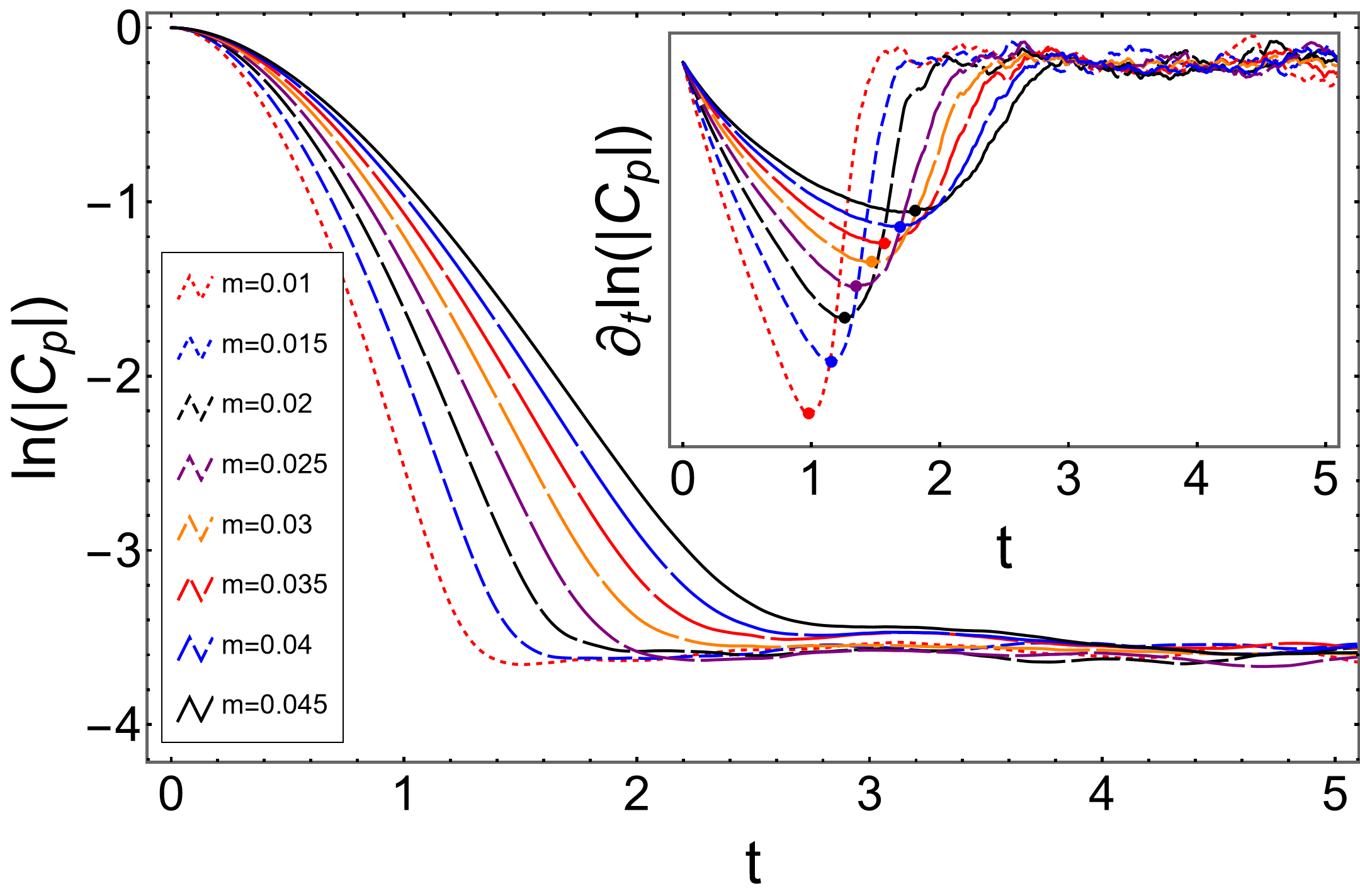}
\caption{\label{fig:PhaseCorrelations} Phase autocorrelation function $C_p(t_1-t_2) = \left<e^{i\ta(t_1,0)-i\ta(t_2,0)}\right>$ in 1 dimension. The logarithm exhibits a quadratic falloff in time, as predicted by eq. 9. The derivative is approximately linear as shown on the inset. The simulation parameters: number of runs $N=1000$, length of system $L = 500$, lattice constant $\Delta L = 1$, $a=10$, $F=0.2$, $D=1$, $B_r=1$, $B_\ta =0$.}
\end{figure}

In Fig. \ref{fig:PhaseCorrelations} we plot the phase autocorrelation function defined in eq. 9 for one dimension. On a logarithmic scale, it exhibits a quadratic falloff as predicted by that equation. The correlations don't decay all the way to zero due to compactness of the phase degree of freedom. In the inset we plot the time derivative of the correlation function, and it exhibits linear behavior for small times, again confirming the prediction. In order to observe the transition from a stretched exponential decay to the usual decay of phase correlations in one dimension $\sim e^{-\sqrt{t}}$, one has to be able to discern the competition between these two regimes before the correlation function reaches the eventual constant.

To see the whether the transition is preserved, we can look for a peak in the response function to an external field. By adding a constant external field $-h_r$ to eq. 13, we can look at:
\begin{equation}
    \begin{split}
        \chi_r = \f{\p \left<r\right>}{\p h_r} \;\;\;,\;\;\; \chi_\om = \f{\p \left<\p_t \ta \right>}{\p h_r} \;.
    \end{split}
\end{equation}

\begin{figure}[!t]
\includegraphics[scale=0.22]{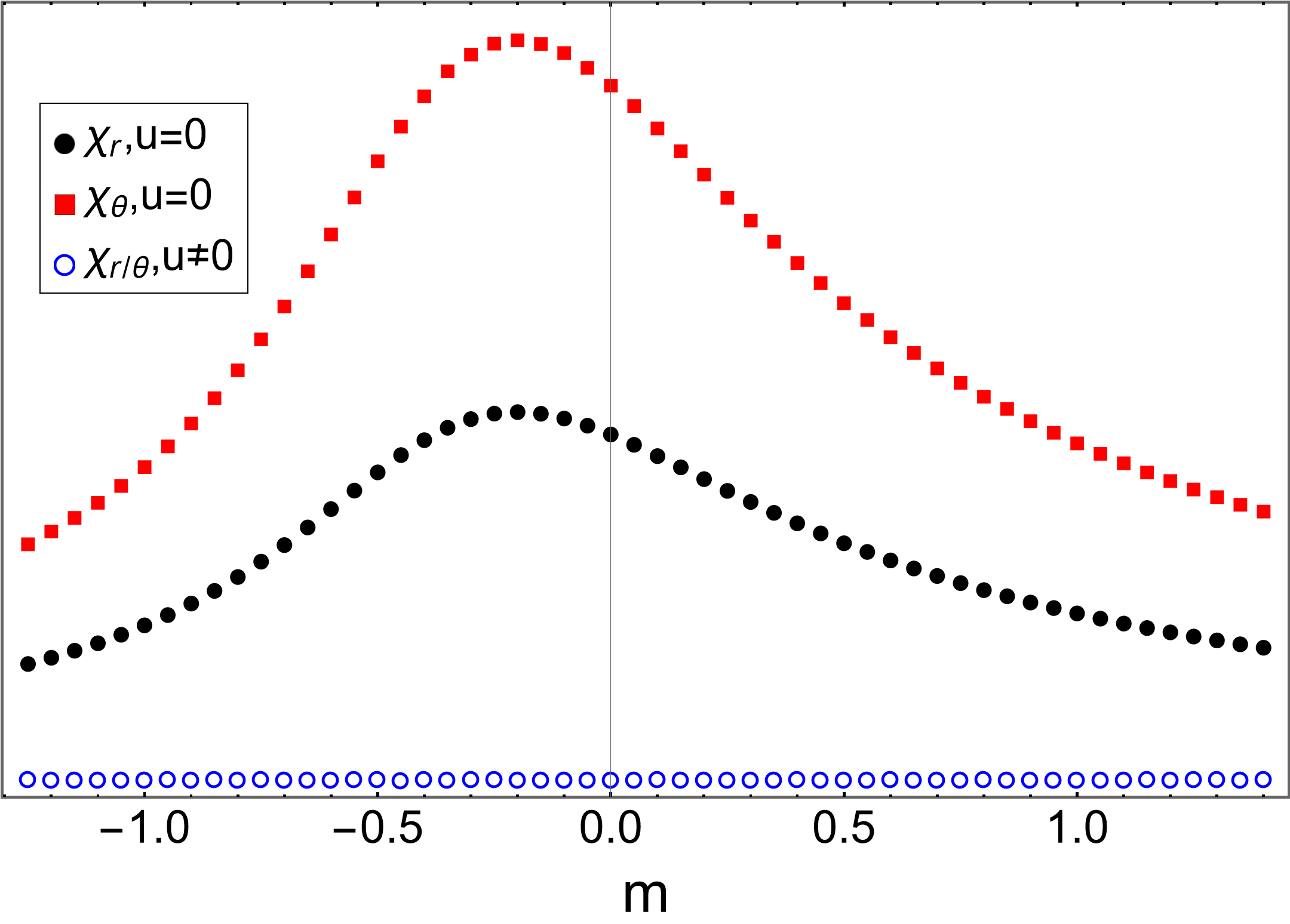}
\caption{\label{fig:radial_response} Response functions for the system in $d=2$ dimensions. All simulations are on a $40 \times 40$ lattice, time step is 0.01, number of runs N=1000. The system runs until it's thermalized. For all simulations $B = 1$, $a=10$, $g=0.5$, $D=1$ and lattice constant is set to $1$. For the frequency response function at $u=0$, we set $F=0.2$. }
\end{figure}

In the thermodynamic limit the response function diverges at the phase transition \cite{Altland2023, Tauber2014}, and will have a peak for a finite system. In Fig. \ref{fig:radial_response} we plot the response functions for both radial d.o.f. and frequency in two dimensions. When $u=0$ we have the model A dynamics for the radius, which exhibits a phase transition in two dimensions. The frequency is proportional to the radial d.o.f., and so the frequency also orders. The peak doesn't necessarily align with the phase transition point exactly, and depends on the size of the system. When $u\neq 0$, the fields exhibit very weak response to external perturbations, signaling that they are in a deep potential well. There is certainly no continuous transition in that case. 

\begin{figure}[!t]
\includegraphics[scale=0.33]{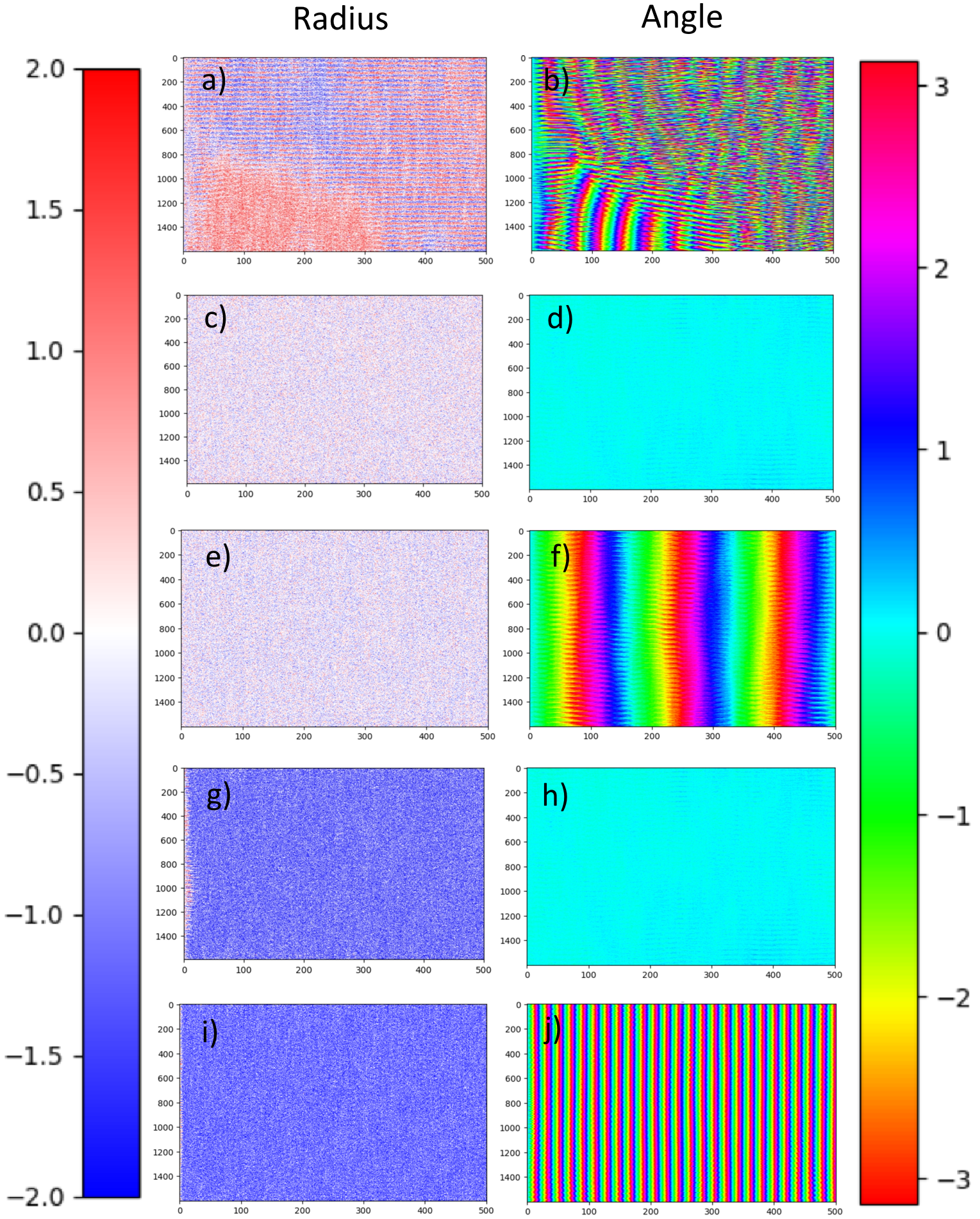}
\caption{\label{fig:ExampleEvolutions} Example evolutions of radius and angle degrees of freedom in $d=2$ dimensions. The grid is stretched to a one-dimensional line for the plot. All simulations are on a $40 \times 40$ lattice, time step is $0.01$, and the time range is $\left[0,500\right]$. The initial condition for the radius is random and the angles are aligned. For all simulations $B = 1$, $a=10$, $g=0.5$, $D=1$ and lattice constant is set to $1$. a),b) $m=-0.7$, $u = 0$, $F=0.3$. c),d) $m=0.5$, $u = -Da$, $F=0$. e),f) $m=0.5$, $u = -Da$, $F=0.3$. g),h) $m=-0.7$, $u = -Da$, $F=0$. i),j) $m=-0.7$, $u = -Da$, $F=0.3$.}
\end{figure}

In Fig. \ref{fig:ExampleEvolutions} we plot example evolutions of both degrees of freedom in a $d=2$ simulation. When $m<0$ and $u=0$ (Fig. \ref{fig:ExampleEvolutions}a) and \ref{fig:ExampleEvolutions}b)) The radial degree of freedom undergoes the usual $\mathcal{Z}_2$ dynamics, forming domains of up and down fields. The angle then rotates according to which configuration the domain is in. When $u \neq 0$, even in the $m>0$ regime, the noise in the angular degree of freedom produces fluctuations which result in a non-zero $\grad \ta$. That produces a downward shift in the radial degree of freedom, Fig. \ref{fig:ExampleEvolutions}c),d),e) and f). When $m<0$, even though there are initial domains of positive $\de r$, they are eventually destroyed by the presence of the $u$ term, Fig. \ref{fig:ExampleEvolutions}g),h). When $F \neq 0$, that destruction is even more apparent and can be understood intuitively. 

\begin{figure}[!t]
\includegraphics[scale=0.33]{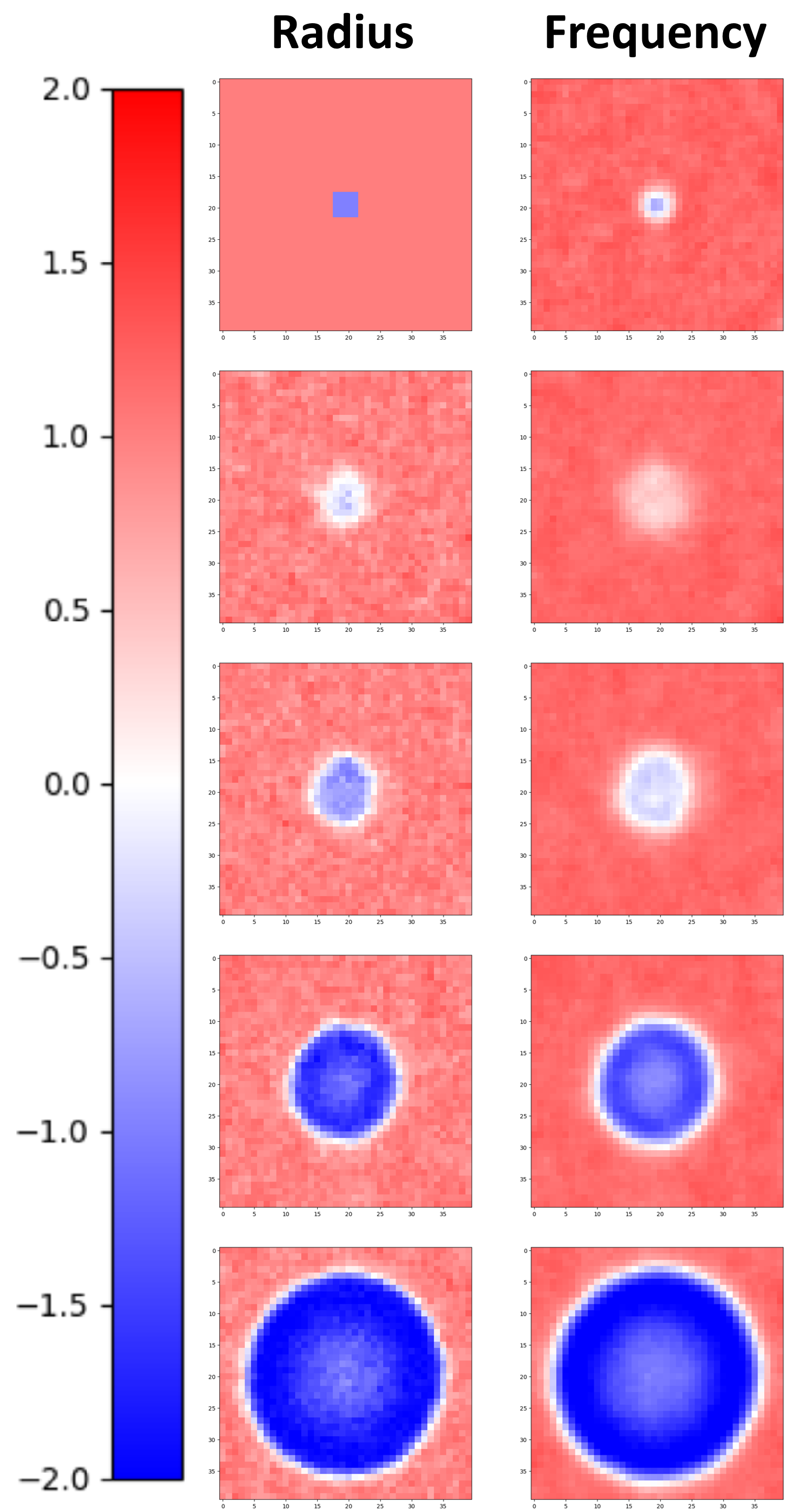}
\caption{\label{fig:ExampleDomain} Example evolutions of radius and frequency of oscillation in $d=2$ dimensions for an initial condition with a small negative domain. The simulation is on a $40 \times 40$ lattice, time step is $0.01$, and the time range is $\left(0,5,7,10,15\right)$ top to bottom. Other parameters are set to $m=-0.5$, $F=0.3$, $B = 0.3$, $a=10$, $g=0.5$, $D=1$ and lattice constant is set to $1$. After a period of accumulating a zero domain wall between the negative and the positive domains, the negative eventually overtakes the positive through a mechanism described in the text.}
\end{figure}

On the edge of two domains, the $F\de r$ term accumulates a phase difference between the two domains. That means that $\grad \ta$ is increasing along the domain wall, which makes the $u$ term larger along the wall. That in turn shifts the radial degree of freedom along the wall, and the negative domains eventually consume the positive through that mechanism. This growth of the domain wall is demonstrated in Fig. \ref{fig:ExampleDomain}. It can also be seen there that as the negative domain overtakes the system, it in fact gets less negative, since the radius orders more, which decreases the size of $\grad \ta$ through the $F$ term. 

Similarly, one can simulate the full field equations that would satisfy all the needed properties. An example is an equation:
\begin{equation}
\begin{split}
    \p_t \psi &= \left[-m\left(|\psi|^2-a^2\right)-g\left(|\psi|^2-a^2\right)^3\right]\psi \\&+ iF\left(|\psi|^2-a^2\right)\psi + D\nabla^2\psi + \xi    
\end{split} \;.
\end{equation}

\begin{figure*}[!t]
\includegraphics[scale=0.4]{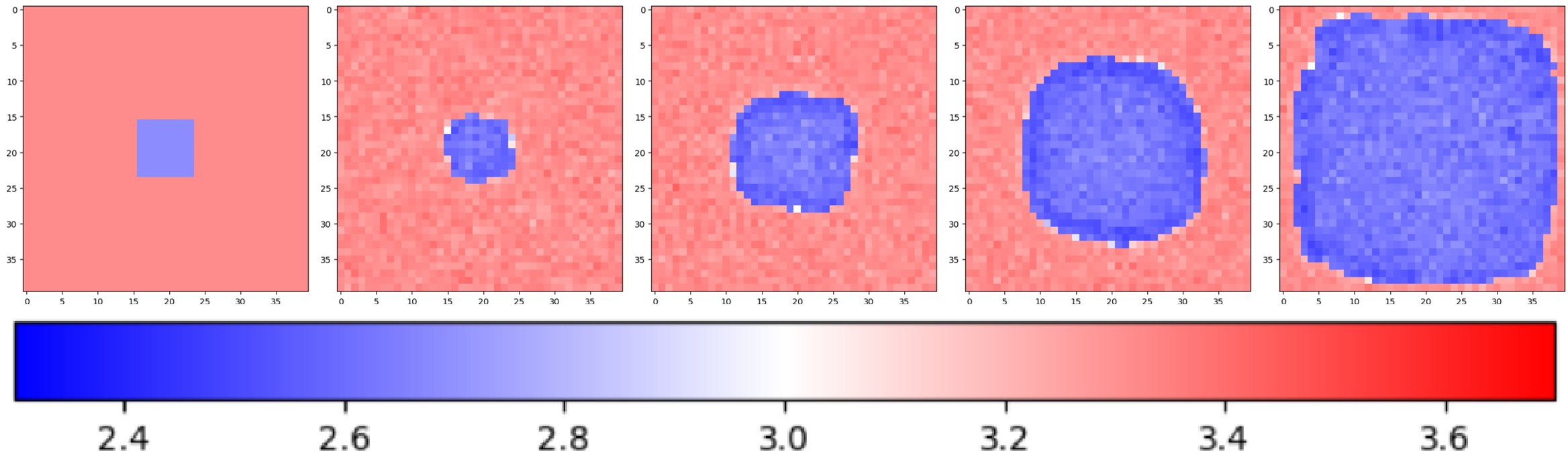}
\caption{\label{fig:FullExampleDomain} Example evolutions of the radius of the full field eq. 22 in $d=2$ dimensions for an initial condition with a small negative domain. The simulation is on a $40 \times 40$ lattice, time step is $0.01$, and the time range is $\left(0,5,15,25,40\right)$ left to right. Other parameters are set to $m=-2$, $F=0.3$, $B = 0.3$, $a=3$, $g=0.5$, $D=1$ and lattice constant is set to $1$. The smaller radius domain eventually overtakes the positive through a mechanism described in the text.}
\end{figure*}

One can check that in if the first line of this equation is seen as a gradient descent in some potential, then that potential will indeed be symmetric around $r=a$ (ignoring the $B/2r$ term, which would only make the lower radius limit cycle less stable). It can also be checked by using the calculation outlined in \cite{Weis2023}, that the transition at $m=0$ is going to be an exceptional point at the mean-field level given $F\neq0$. When these equations are simulated in the two-cycle regime, we again see the same mechanism where a domain in a smaller radius state grows and consumes the entire system, see Fig. \ref{fig:FullExampleDomain}.

All of the simulations presented above are qualitative demonstrations of the analytical claims made in the previous sections. Just as a detailed perturbative or renormalization group analysis is likely necessary to make more precise analytic claims, more detailed numerical approaches can also be taken to study this system. Those could include finite-size scaling techniques, larger lattice sizes, Binder coefficient as a measure of the transition, e.t.c. We leave both sides of the problem for future work.

\section*{\label{sec:conclusion}Conclusion}
Limit cycles are a phenomenon that is present in a variety of systems and models, ranging from physics to social sciences. They are one of the most elementary steady states that a system driven out of equilibrium can settle into. Their ubiquity and relative simplicity makes them a prime candidate for exploring non-equilibrium phases of matter from a field theoretical perspective. The phenomenology of limit cycles includes the possibility of birhythmicity - a regime of a system where it supports two stable limit cycles. Birhythmicity can be found in a variety of disciplines and applications as well, like biochemistry, biology, modeling of social behavior, and others. Through a generalized complex Ginzburg-Landau equation, we introduced a class of models with rotational symmetry that can be used to describe transitions to birhythmicity.

As a generic feature we also included a coupling between the magnitude of the order parameter and the frequency of it's oscillations. Such terms often arise in real systems, and have major effects, for example for the functioning of the brain. This feature drives the system out of equilibrium and has profound consequences on it's behavior. 

In order to explore a continuous onset of birhythmicity, where a single stable limit cycle splits into two stable ones, we used a pitchfork bifurcation around a finite magnitude of the complex field variable. This choice was made in order to draw comparison to the equilibrium second-order transition of a scalar field. With that, we described the lowest order field-theoretic action resulting from such a model. 

First, we show that at a Gaussian level, the coupling between phase and radius leads to a stretched exponential decay of phase correlations below certain length and time scales. We also demonstrate that that coupling can be a source of $1/f$ noise for the phase.

Then, if the transition is indeed continuous, it happens through a critical exceptional point (CEP). Exceptional points (EPs) and CEPs are special points in parameter space that are responsible for a variety of unusual behavior and statistical properties of the system. They are a focus of many studies in their own right, and seem to be an essential feature to be contended with when non-equilibrium models are considered. For our model, the statistical properties of the phase correlations are altered drastically at the CEP, and the fluctuations are enhanced. That is a known feature of exceptional points, and hints at the appearance of new universality classes as far as the transition is concerned. We demonstrated how a coupling between the phase and the radius in such a transition would lead to a critical exceptional point.

We show that although we start with a model of a continuous transition, both the diffusion and the fact that it occurs around a finite magnitude inevitably leads to symmetry breaking terms that should erase it. One of such terms produces an "external field" as far as the effective action of the radial degree of freedom is concerned. In that case, the birhythmicity onsets through a second stable limit cycle occurring through a saddle-node bifurcation, which is a pathway to birhythmicity often observed in real systems and simulations. However, the symmetry breaking term can be irrelevant in the renormalization group sense, which we show to be the case at the scaling level for dimensions less than two. Other terms have the opposite behavior at the scaling level. Although we could arrive at an Ising-like model for the radial transition in one dimension through an RG flow, the domain walls will destroy the order anyway. We leave the more detailed exploration of this problem for future work.

We also argue that a range of other terms are inevitably going to be produced by a renormalization group flow. As an example we show that the non-equilibrium coupling induces KPZ dynamics for the phase degree of freedom. That is also a generic feature of the phase that is coupled to amplitude fluctuations. Here we demonstrate it using a field-theoretic language. While this effect breaks the $Z_2$ symmetry and therefore breaks the continuous symmetry of the phase transition, it is possible that parameter-tuning can restore a continuous critical point by a judicious cancellation of terms; that certainly can happen at the mean-field level \cite{Zelle2023, Daviet2024, Daviet2024b, huang25} 

Finally, we demonstrated the analytic predictions using lattice simulations in low dimensions. Most importantly, we confirm that such a transition is inevitably destroyed by the presence of non-linearities appearing from diffusion. We do so both by simulating the approximated dynamics developed in the text, as well as for full field dynamics of an example Ginzburg-Landau equation. We also describe the intuitive mechanism of how droplets of negative domains eventually grow to dominate the entire system. 

Although we approached this study from an abstract perspective, we believe it should provide a suitable framework for exploring a wide range of problems. First, as discussed in the Introduction, there is a variety of real systems exhibiting birhythmicity (and transitions to birhythmicity). Spatial coupling is a natural ingredient that can be added to those, providing a basis for using dynamical field theory. More broadly, assuming the ideas of universality are applicable to such non-equilibrium systems, our work should provide a basis for studying any system within the appropriate universality class. Our model is an example of the type of field theoretic descriptions that one can construct when considering non-equilibrium systems. Such a description allows for the use of field-theoretic tools in order to answer the prototypical questions of statistical physics. In this case, we demonstrated a number of properties of the model of birhythmicity, leaving a more detailed analysis for the universality class of the transition for future work. 

\section*{Acknowledements}
The authors would like to thank Sebastian Diehl and Ryo Hanai for helpful discussions. This research was partly supported by the National Science Foundation through the Center for Living Systems (NSF-PHY 2317138) and by the NSF-MPS-PHY award 2207383. 

\nocite{*}
\bibliography{apssamp}

\providecommand{\noopsort}[1]{}\providecommand{\singleletter}[1]{#1}%
\begin{thebibliography}{60}%
\makeatletter
\providecommand \@ifxundefined [1]{%
 \@ifx{#1\undefined}
}%
\providecommand \@ifnum [1]{%
 \ifnum #1\expandafter \@firstoftwo
 \else \expandafter \@secondoftwo
 \fi
}%
\providecommand \@ifx [1]{%
 \ifx #1\expandafter \@firstoftwo
 \else \expandafter \@secondoftwo
 \fi
}%
\providecommand \natexlab [1]{#1}%
\providecommand \enquote  [1]{``#1''}%
\providecommand \bibnamefont  [1]{#1}%
\providecommand \bibfnamefont [1]{#1}%
\providecommand \citenamefont [1]{#1}%
\providecommand \href@noop [0]{\@secondoftwo}%
\providecommand \href [0]{\begingroup \@sanitize@url \@href}%
\providecommand \@href[1]{\@@startlink{#1}\@@href}%
\providecommand \@@href[1]{\endgroup#1\@@endlink}%
\providecommand \@sanitize@url [0]{\catcode `\\12\catcode `\$12\catcode `\&12\catcode `\#12\catcode `\^12\catcode `\_12\catcode `\%12\relax}%
\providecommand \@@startlink[1]{}%
\providecommand \@@endlink[0]{}%
\providecommand \url  [0]{\begingroup\@sanitize@url \@url }%
\providecommand \@url [1]{\endgroup\@href {#1}{\urlprefix }}%
\providecommand \urlprefix  [0]{URL }%
\providecommand \Eprint [0]{\href }%
\providecommand \doibase [0]{https://doi.org/}%
\providecommand \selectlanguage [0]{\@gobble}%
\providecommand \bibinfo  [0]{\@secondoftwo}%
\providecommand \bibfield  [0]{\@secondoftwo}%
\providecommand \translation [1]{[#1]}%
\providecommand \BibitemOpen [0]{}%
\providecommand \bibitemStop [0]{}%
\providecommand \bibitemNoStop [0]{.\EOS\space}%
\providecommand \EOS [0]{\spacefactor3000\relax}%
\providecommand \BibitemShut  [1]{\csname bibitem#1\endcsname}%
\let\auto@bib@innerbib\@empty
\bibitem [{\citenamefont {Altland}\ and\ \citenamefont {Simons}(2023)}]{Altland2023}%
  \BibitemOpen
  \bibfield  {author} {\bibinfo {author} {\bibfnamefont {A.}~\bibnamefont {Altland}}\ and\ \bibinfo {author} {\bibfnamefont {B.}~\bibnamefont {Simons}},\ }\href@noop {} {\emph {\bibinfo {title} {Condensed matter field theory}}}\ (\bibinfo  {publisher} {Cambridge University Press},\ \bibinfo {year} {2023})\BibitemShut {NoStop}%
\bibitem [{\citenamefont {Tauber}(2014)}]{Tauber2014}%
  \BibitemOpen
  \bibfield  {author} {\bibinfo {author} {\bibfnamefont {U.~C.}\ \bibnamefont {Tauber}},\ }\href@noop {} {\emph {\bibinfo {title} {Critical dynamics: A field theory approach to equilibrium and non-equilibrium scaling behavior}}}\ (\bibinfo  {publisher} {Cambridge University Press},\ \bibinfo {year} {2014})\BibitemShut {NoStop}%
\bibitem [{\citenamefont {Kamenev}(2023)}]{Kamenev2023}%
  \BibitemOpen
  \bibfield  {author} {\bibinfo {author} {\bibfnamefont {A.}~\bibnamefont {Kamenev}},\ }\href@noop {} {\emph {\bibinfo {title} {Field theory of non-equilibrium systems}}}\ (\bibinfo  {publisher} {Cambridge University Press},\ \bibinfo {year} {2023})\BibitemShut {NoStop}%
\bibitem [{\citenamefont {Cross}\ and\ \citenamefont {Hohenberg}(1993)}]{Cross1993}%
  \BibitemOpen
  \bibfield  {author} {\bibinfo {author} {\bibfnamefont {M.~C.}\ \bibnamefont {Cross}}\ and\ \bibinfo {author} {\bibfnamefont {P.~C.}\ \bibnamefont {Hohenberg}},\ }\bibfield  {title} {\bibinfo {title} {Pattern formation outside of equilibrium},\ }\href {https://doi.org/10.1103/revmodphys.65.851} {\bibfield  {journal} {\bibinfo  {journal} {Rev. Mod. Phys.}\ }\textbf {\bibinfo {volume} {65}},\ \bibinfo {pages} {851–1112} (\bibinfo {year} {1993})}\BibitemShut {NoStop}%
\bibitem [{\citenamefont {Zelle}\ \emph {et~al.}(2024)\citenamefont {Zelle}, \citenamefont {Daviet}, \citenamefont {Rosch},\ and\ \citenamefont {Diehl}}]{Zelle2023}%
  \BibitemOpen
  \bibfield  {author} {\bibinfo {author} {\bibfnamefont {C.~P.}\ \bibnamefont {Zelle}}, \bibinfo {author} {\bibfnamefont {R.}~\bibnamefont {Daviet}}, \bibinfo {author} {\bibfnamefont {A.}~\bibnamefont {Rosch}},\ and\ \bibinfo {author} {\bibfnamefont {S.}~\bibnamefont {Diehl}},\ }\bibfield  {title} {\bibinfo {title} {Universal phenomenology at critical exceptional points of nonequilibrium o(n) models},\ }\bibfield  {journal} {\bibinfo  {journal} {Phys. Rev. X}\ }\textbf {\bibinfo {volume} {14}},\ \href {https://doi.org/10.1103/physrevx.14.021052} {10.1103/physrevx.14.021052} (\bibinfo {year} {2024})\BibitemShut {NoStop}%
\bibitem [{\citenamefont {Daviet}\ \emph {et~al.}(2024{\natexlab{a}})\citenamefont {Daviet}, \citenamefont {Zelle}, \citenamefont {Rosch},\ and\ \citenamefont {Diehl}}]{Daviet2024}%
  \BibitemOpen
  \bibfield  {author} {\bibinfo {author} {\bibfnamefont {R.}~\bibnamefont {Daviet}}, \bibinfo {author} {\bibfnamefont {C.~P.}\ \bibnamefont {Zelle}}, \bibinfo {author} {\bibfnamefont {A.}~\bibnamefont {Rosch}},\ and\ \bibinfo {author} {\bibfnamefont {S.}~\bibnamefont {Diehl}},\ }\bibfield  {title} {\bibinfo {title} {Nonequilibrium criticality at the onset of time-crystalline order},\ }\bibfield  {journal} {\bibinfo  {journal} {Phys. Rev. Lett.}\ }\textbf {\bibinfo {volume} {132}},\ \href {https://doi.org/10.1103/physrevlett.132.167102} {10.1103/physrevlett.132.167102} (\bibinfo {year} {2024}{\natexlab{a}})\BibitemShut {NoStop}%
\bibitem [{\citenamefont {Strogatz}(2024)}]{Strogatz2024}%
  \BibitemOpen
  \bibfield  {author} {\bibinfo {author} {\bibfnamefont {S.}~\bibnamefont {Strogatz}},\ }\href@noop {} {\emph {\bibinfo {title} {Nonlinear Dynamics and Chaos: With applications to physics, biology, chemistry, and engineering}}}\ (\bibinfo  {publisher} {CRC Press},\ \bibinfo {year} {2024})\BibitemShut {NoStop}%
\bibitem [{\citenamefont {Guckenheimer}\ \emph {et~al.}(2013)\citenamefont {Guckenheimer}, \citenamefont {Holmes},\ and\ \citenamefont {Sirovich}}]{Guckenheimer2013}%
  \BibitemOpen
  \bibfield  {author} {\bibinfo {author} {\bibfnamefont {J.}~\bibnamefont {Guckenheimer}}, \bibinfo {author} {\bibfnamefont {P.}~\bibnamefont {Holmes}},\ and\ \bibinfo {author} {\bibfnamefont {L.}~\bibnamefont {Sirovich}},\ }\href@noop {} {\emph {\bibinfo {title} {Nonlinear oscillations, dynamical systems, and bifurcations of Vector Fields}}}\ (\bibinfo  {publisher} {Springer},\ \bibinfo {year} {2013})\BibitemShut {NoStop}%
\bibitem [{\citenamefont {Kuramoto}(2003)}]{Kuramoto2003}%
  \BibitemOpen
  \bibfield  {author} {\bibinfo {author} {\bibfnamefont {Y.}~\bibnamefont {Kuramoto}},\ }\href@noop {} {\emph {\bibinfo {title} {Chemical oscillations, waves, and turbulence}}}\ (\bibinfo  {publisher} {Dover Publications},\ \bibinfo {year} {2003})\BibitemShut {NoStop}%
\bibitem [{\citenamefont {Pikovskij}\ \emph {et~al.}(2007)\citenamefont {Pikovskij}, \citenamefont {Rosenblum},\ and\ \citenamefont {Kurths}}]{Pikovskij2007}%
  \BibitemOpen
  \bibfield  {author} {\bibinfo {author} {\bibfnamefont {A.~S.}\ \bibnamefont {Pikovskij}}, \bibinfo {author} {\bibfnamefont {M.}~\bibnamefont {Rosenblum}},\ and\ \bibinfo {author} {\bibfnamefont {J.}~\bibnamefont {Kurths}},\ }\href@noop {} {\emph {\bibinfo {title} {Synchronization: A universal concept in Nonlinear Sciences}}}\ (\bibinfo  {publisher} {Cambridge Univ. Press},\ \bibinfo {year} {2007})\BibitemShut {NoStop}%
\bibitem [{\citenamefont {Tian}\ \emph {et~al.}(2022)\citenamefont {Tian}, \citenamefont {Tan}, \citenamefont {Hou}, \citenamefont {Li}, \citenamefont {Cheng}, \citenamefont {Qiu}, \citenamefont {Weng}, \citenamefont {Chen},\ and\ \citenamefont {Sun}}]{Tian2022}%
  \BibitemOpen
  \bibfield  {author} {\bibinfo {author} {\bibfnamefont {Y.}~\bibnamefont {Tian}}, \bibinfo {author} {\bibfnamefont {Z.}~\bibnamefont {Tan}}, \bibinfo {author} {\bibfnamefont {H.}~\bibnamefont {Hou}}, \bibinfo {author} {\bibfnamefont {G.}~\bibnamefont {Li}}, \bibinfo {author} {\bibfnamefont {A.}~\bibnamefont {Cheng}}, \bibinfo {author} {\bibfnamefont {Y.}~\bibnamefont {Qiu}}, \bibinfo {author} {\bibfnamefont {K.}~\bibnamefont {Weng}}, \bibinfo {author} {\bibfnamefont {C.}~\bibnamefont {Chen}},\ and\ \bibinfo {author} {\bibfnamefont {P.}~\bibnamefont {Sun}},\ }\bibfield  {title} {\bibinfo {title} {Theoretical foundations of studying criticality in the brain},\ }\href {https://doi.org/10.1162/netn_a_00269} {\bibfield  {journal} {\bibinfo  {journal} {Netw. Neurosci.}\ }\textbf {\bibinfo {volume} {6}},\ \bibinfo {pages} {1148–1185} (\bibinfo {year} {2022})}\BibitemShut {NoStop}%
\bibitem [{\citenamefont {Goldstone}\ and\ \citenamefont {Garmire}(1984)}]{Goldstone1984}%
  \BibitemOpen
  \bibfield  {author} {\bibinfo {author} {\bibfnamefont {J.~A.}\ \bibnamefont {Goldstone}}\ and\ \bibinfo {author} {\bibfnamefont {E.}~\bibnamefont {Garmire}},\ }\bibfield  {title} {\bibinfo {title} {Intrinsic optical bistability in nonlinear media},\ }\href {https://doi.org/10.1103/physrevlett.53.910} {\bibfield  {journal} {\bibinfo  {journal} {Phys. Rev. Lett.}\ }\textbf {\bibinfo {volume} {53}},\ \bibinfo {pages} {910–913} (\bibinfo {year} {1984})}\BibitemShut {NoStop}%
\bibitem [{\citenamefont {Ritter}\ and\ \citenamefont {Haug}(1993)}]{Ritter1993}%
  \BibitemOpen
  \bibfield  {author} {\bibinfo {author} {\bibfnamefont {A.}~\bibnamefont {Ritter}}\ and\ \bibinfo {author} {\bibfnamefont {H.}~\bibnamefont {Haug}},\ }\bibfield  {title} {\bibinfo {title} {Theory of the bistable limit cycle behavior of laser diodes induced by weak optical feedback},\ }\href {https://doi.org/10.1109/3.214491} {\bibfield  {journal} {\bibinfo  {journal} {IEEE J. Quantum Electron.}\ }\textbf {\bibinfo {volume} {29}},\ \bibinfo {pages} {1064–1070} (\bibinfo {year} {1993})}\BibitemShut {NoStop}%
\bibitem [{\citenamefont {Biswas}\ \emph {et~al.}(2016)\citenamefont {Biswas}, \citenamefont {Banerjee},\ and\ \citenamefont {Kurths}}]{Biswas2016}%
  \BibitemOpen
  \bibfield  {author} {\bibinfo {author} {\bibfnamefont {D.}~\bibnamefont {Biswas}}, \bibinfo {author} {\bibfnamefont {T.}~\bibnamefont {Banerjee}},\ and\ \bibinfo {author} {\bibfnamefont {J.}~\bibnamefont {Kurths}},\ }\bibfield  {title} {\bibinfo {title} {Control of birhythmicity through conjugate self-feedback: Theory and experiment},\ }\bibfield  {journal} {\bibinfo  {journal} {Phys. Rev. E}\ }\textbf {\bibinfo {volume} {94}},\ \href {https://doi.org/10.1103/physreve.94.042226} {10.1103/physreve.94.042226} (\bibinfo {year} {2016})\BibitemShut {NoStop}%
\bibitem [{\citenamefont {Goldbeter}(2018)}]{Goldbeter2018}%
  \BibitemOpen
  \bibfield  {author} {\bibinfo {author} {\bibfnamefont {A.}~\bibnamefont {Goldbeter}},\ }\bibfield  {title} {\bibinfo {title} {Dissipative structures in biological systems: Bistability, oscillations, spatial patterns and waves},\ }\href {https://doi.org/10.1098/rsta.2017.0376} {\bibfield  {journal} {\bibinfo  {journal} {Philos. Trans. A Math. Phys. Eng. Sci.}\ }\textbf {\bibinfo {volume} {376}},\ \bibinfo {pages} {20170376} (\bibinfo {year} {2018})}\BibitemShut {NoStop}%
\bibitem [{\citenamefont {Decroly}\ and\ \citenamefont {Goldbeter}(1982)}]{Decroly1982}%
  \BibitemOpen
  \bibfield  {author} {\bibinfo {author} {\bibfnamefont {O.}~\bibnamefont {Decroly}}\ and\ \bibinfo {author} {\bibfnamefont {A.}~\bibnamefont {Goldbeter}},\ }\bibfield  {title} {\bibinfo {title} {Birhythmicity, chaos, and other patterns of temporal self-organization in a multiply regulated biochemical system.},\ }\href {https://doi.org/10.1073/pnas.79.22.6917} {\bibfield  {journal} {\bibinfo  {journal} {Proc. Natl. Acad. Sci.}\ }\textbf {\bibinfo {volume} {79}},\ \bibinfo {pages} {6917–6921} (\bibinfo {year} {1982})}\BibitemShut {NoStop}%
\bibitem [{\citenamefont {Tsumoto}\ \emph {et~al.}(2006)\citenamefont {Tsumoto}, \citenamefont {Yoshinaga}, \citenamefont {Iida}, \citenamefont {Kawakami},\ and\ \citenamefont {Aihara}}]{Tsumoto2006}%
  \BibitemOpen
  \bibfield  {author} {\bibinfo {author} {\bibfnamefont {K.}~\bibnamefont {Tsumoto}}, \bibinfo {author} {\bibfnamefont {T.}~\bibnamefont {Yoshinaga}}, \bibinfo {author} {\bibfnamefont {H.}~\bibnamefont {Iida}}, \bibinfo {author} {\bibfnamefont {H.}~\bibnamefont {Kawakami}},\ and\ \bibinfo {author} {\bibfnamefont {K.}~\bibnamefont {Aihara}},\ }\bibfield  {title} {\bibinfo {title} {Bifurcations in a mathematical model for circadian oscillations of clock genes},\ }\href {https://doi.org/10.1016/j.jtbi.2005.07.017} {\bibfield  {journal} {\bibinfo  {journal} {J. Theor. Biol.}\ }\textbf {\bibinfo {volume} {239}},\ \bibinfo {pages} {101–122} (\bibinfo {year} {2006})}\BibitemShut {NoStop}%
\bibitem [{\citenamefont {De~la Fuente}(1999)}]{Fuente1999}%
  \BibitemOpen
  \bibfield  {author} {\bibinfo {author} {\bibfnamefont {I.}~\bibnamefont {De~la Fuente}},\ }\bibfield  {title} {\bibinfo {title} {Diversity of temporal self-organized behaviors in a biochemical system},\ }\href {https://doi.org/10.1016/s0303-2647(98)00094-x} {\bibfield  {journal} {\bibinfo  {journal} {Biosystems}\ }\textbf {\bibinfo {volume} {50}},\ \bibinfo {pages} {83–97} (\bibinfo {year} {1999})}\BibitemShut {NoStop}%
\bibitem [{\citenamefont {Yamapi}\ \emph {et~al.}(2010)\citenamefont {Yamapi}, \citenamefont {Filatrella},\ and\ \citenamefont {Aziz-Alaoui}}]{Yamapi2010}%
  \BibitemOpen
  \bibfield  {author} {\bibinfo {author} {\bibfnamefont {R.}~\bibnamefont {Yamapi}}, \bibinfo {author} {\bibfnamefont {G.}~\bibnamefont {Filatrella}},\ and\ \bibinfo {author} {\bibfnamefont {M.~A.}\ \bibnamefont {Aziz-Alaoui}},\ }\bibfield  {title} {\bibinfo {title} {Global stability analysis of birhythmicity in a self-sustained oscillator},\ }\bibfield  {journal} {\bibinfo  {journal} {Chaos}\ }\textbf {\bibinfo {volume} {20}},\ \href {https://doi.org/10.1063/1.3309014} {10.1063/1.3309014} (\bibinfo {year} {2010})\BibitemShut {NoStop}%
\bibitem [{\citenamefont {Kar}\ and\ \citenamefont {Ray}(2004)}]{Kar2004}%
  \BibitemOpen
  \bibfield  {author} {\bibinfo {author} {\bibfnamefont {S.}~\bibnamefont {Kar}}\ and\ \bibinfo {author} {\bibfnamefont {D.~S.}\ \bibnamefont {Ray}},\ }\bibfield  {title} {\bibinfo {title} {Large fluctuations and nonlinear dynamics of birhythmicity},\ }\href {https://doi.org/10.1209/epl/i2003-10277-9} {\bibfield  {journal} {\bibinfo  {journal} {Europhys. Lett.}\ }\textbf {\bibinfo {volume} {67}},\ \bibinfo {pages} {137–143} (\bibinfo {year} {2004})}\BibitemShut {NoStop}%
\bibitem [{\citenamefont {Cook}\ \emph {et~al.}(2022)\citenamefont {Cook}, \citenamefont {Peterson}, \citenamefont {Woldman},\ and\ \citenamefont {Terry}}]{Cook2022}%
  \BibitemOpen
  \bibfield  {author} {\bibinfo {author} {\bibfnamefont {B.~J.}\ \bibnamefont {Cook}}, \bibinfo {author} {\bibfnamefont {A.~D.}\ \bibnamefont {Peterson}}, \bibinfo {author} {\bibfnamefont {W.}~\bibnamefont {Woldman}},\ and\ \bibinfo {author} {\bibfnamefont {J.~R.}\ \bibnamefont {Terry}},\ }\bibfield  {title} {\bibinfo {title} {Neural field models: A mathematical overview and unifying framework},\ }\bibfield  {journal} {\bibinfo  {journal} {Math. Neurosci. Appl.}\ }\textbf {\bibinfo {volume} {Volume 2}},\ \href {https://doi.org/10.46298/mna.7284} {10.46298/mna.7284} (\bibinfo {year} {2022})\BibitemShut {NoStop}%
\bibitem [{\citenamefont {Basar}(2013)}]{Basar2013}%
  \BibitemOpen
  \bibfield  {author} {\bibinfo {author} {\bibfnamefont {E.}~\bibnamefont {Basar}},\ }\bibfield  {title} {\bibinfo {title} {Brain oscillations in neuropsychiatric disease},\ }\href {https://doi.org/10.31887/dcns.2013.15.3/ebasar} {\bibfield  {journal} {\bibinfo  {journal} {Dialogues Clin. Neurosci.}\ }\textbf {\bibinfo {volume} {15}},\ \bibinfo {pages} {291–300} (\bibinfo {year} {2013})}\BibitemShut {NoStop}%
\bibitem [{\citenamefont {di~Santo}\ \emph {et~al.}(2018)\citenamefont {di~Santo}, \citenamefont {Villegas}, \citenamefont {Burioni},\ and\ \citenamefont {Muñoz}}]{diSanto2018}%
  \BibitemOpen
  \bibfield  {author} {\bibinfo {author} {\bibfnamefont {S.}~\bibnamefont {di~Santo}}, \bibinfo {author} {\bibfnamefont {P.}~\bibnamefont {Villegas}}, \bibinfo {author} {\bibfnamefont {R.}~\bibnamefont {Burioni}},\ and\ \bibinfo {author} {\bibfnamefont {M.~A.}\ \bibnamefont {Muñoz}},\ }\bibfield  {title} {\bibinfo {title} {Landau–ginzburg theory of cortex dynamics: Scale-free avalanches emerge at the edge of synchronization},\ }\bibfield  {journal} {\bibinfo  {journal} {Proc. Natl. Acad. Sci.}\ }\textbf {\bibinfo {volume} {115}},\ \href {https://doi.org/10.1073/pnas.1712989115} {10.1073/pnas.1712989115} (\bibinfo {year} {2018})\BibitemShut {NoStop}%
\bibitem [{\citenamefont {Schroeder}\ and\ \citenamefont {Lakatos}(2009)}]{Schroeder2009}%
  \BibitemOpen
  \bibfield  {author} {\bibinfo {author} {\bibfnamefont {C.~E.}\ \bibnamefont {Schroeder}}\ and\ \bibinfo {author} {\bibfnamefont {P.}~\bibnamefont {Lakatos}},\ }\bibfield  {title} {\bibinfo {title} {Low-frequency neuronal oscillations as instruments of sensory selection},\ }\href {https://doi.org/10.1016/j.tins.2008.09.012} {\bibfield  {journal} {\bibinfo  {journal} {Trends Neurosci.}\ }\textbf {\bibinfo {volume} {32}},\ \bibinfo {pages} {9–18} (\bibinfo {year} {2009})}\BibitemShut {NoStop}%
\bibitem [{\citenamefont {Slepukhina}\ \emph {et~al.}(2023)\citenamefont {Slepukhina}, \citenamefont {Bashkirtseva}, \citenamefont {Kügler},\ and\ \citenamefont {Ryashko}}]{Slepukhina2023}%
  \BibitemOpen
  \bibfield  {author} {\bibinfo {author} {\bibfnamefont {E.}~\bibnamefont {Slepukhina}}, \bibinfo {author} {\bibfnamefont {I.}~\bibnamefont {Bashkirtseva}}, \bibinfo {author} {\bibfnamefont {P.}~\bibnamefont {Kügler}},\ and\ \bibinfo {author} {\bibfnamefont {L.}~\bibnamefont {Ryashko}},\ }\bibfield  {title} {\bibinfo {title} {Noise-driven bursting birhythmicity in the hindmarsh–rose neuron model},\ }\bibfield  {journal} {\bibinfo  {journal} {Chaos}\ }\textbf {\bibinfo {volume} {33}},\ \href {https://doi.org/10.1063/5.0134561} {10.1063/5.0134561} (\bibinfo {year} {2023})\BibitemShut {NoStop}%
\bibitem [{\citenamefont {Wilson}\ and\ \citenamefont {Cowan}(1972)}]{Wilson1972}%
  \BibitemOpen
  \bibfield  {author} {\bibinfo {author} {\bibfnamefont {H.~R.}\ \bibnamefont {Wilson}}\ and\ \bibinfo {author} {\bibfnamefont {J.~D.}\ \bibnamefont {Cowan}},\ }\bibfield  {title} {\bibinfo {title} {Excitatory and inhibitory interactions in localized populations of model neurons},\ }\href {https://doi.org/10.1016/s0006-3495(72)86068-5} {\bibfield  {journal} {\bibinfo  {journal} {Biophys. J.}\ }\textbf {\bibinfo {volume} {12}},\ \bibinfo {pages} {1–24} (\bibinfo {year} {1972})}\BibitemShut {NoStop}%
\bibitem [{\citenamefont {Milton}(2010)}]{Milton2010}%
  \BibitemOpen
  \bibfield  {author} {\bibinfo {author} {\bibfnamefont {J.~G.}\ \bibnamefont {Milton}},\ }\bibfield  {title} {\bibinfo {title} {Epilepsy as a dynamic disease: A tutorial of the past with an eye to the future},\ }\href {https://doi.org/10.1016/j.yebeh.2010.03.002} {\bibfield  {journal} {\bibinfo  {journal} {Epilepsy Behav.}\ }\textbf {\bibinfo {volume} {18}},\ \bibinfo {pages} {33–44} (\bibinfo {year} {2010})}\BibitemShut {NoStop}%
\bibitem [{\citenamefont {Junges}\ \emph {et~al.}(2020)\citenamefont {Junges}, \citenamefont {Woldman}, \citenamefont {Benjamin},\ and\ \citenamefont {Terry}}]{Junges2020}%
  \BibitemOpen
  \bibfield  {author} {\bibinfo {author} {\bibfnamefont {L.}~\bibnamefont {Junges}}, \bibinfo {author} {\bibfnamefont {W.}~\bibnamefont {Woldman}}, \bibinfo {author} {\bibfnamefont {O.~J.}\ \bibnamefont {Benjamin}},\ and\ \bibinfo {author} {\bibfnamefont {J.~R.}\ \bibnamefont {Terry}},\ }\bibfield  {title} {\bibinfo {title} {Epilepsy surgery: Evaluating robustness using dynamic network models},\ }\bibfield  {journal} {\bibinfo  {journal} {Chaos}\ }\textbf {\bibinfo {volume} {30}},\ \href {https://doi.org/10.1063/5.0022171} {10.1063/5.0022171} (\bibinfo {year} {2020})\BibitemShut {NoStop}%
\bibitem [{\citenamefont {Harrington}\ \emph {et~al.}(2024)\citenamefont {Harrington}, \citenamefont {Kissack}, \citenamefont {Terry}, \citenamefont {Woldman},\ and\ \citenamefont {Junges}}]{Harrington2024}%
  \BibitemOpen
  \bibfield  {author} {\bibinfo {author} {\bibfnamefont {E.~G.}\ \bibnamefont {Harrington}}, \bibinfo {author} {\bibfnamefont {P.}~\bibnamefont {Kissack}}, \bibinfo {author} {\bibfnamefont {J.~R.}\ \bibnamefont {Terry}}, \bibinfo {author} {\bibfnamefont {W.}~\bibnamefont {Woldman}},\ and\ \bibinfo {author} {\bibfnamefont {L.}~\bibnamefont {Junges}},\ }\bibfield  {title} {\bibinfo {title} {Treatment effects in epilepsy: A mathematical framework for understanding response over time},\ }\bibfield  {journal} {\bibinfo  {journal} {Front. Netw. Physiol.}\ }\textbf {\bibinfo {volume} {4}},\ \href {https://doi.org/10.3389/fnetp.2024.1308501} {10.3389/fnetp.2024.1308501} (\bibinfo {year} {2024})\BibitemShut {NoStop}%
\bibitem [{\citenamefont {Acebrón}\ \emph {et~al.}(2005)\citenamefont {Acebrón}, \citenamefont {Bonilla}, \citenamefont {Pérez~Vicente}, \citenamefont {Ritort},\ and\ \citenamefont {Spigler}}]{Acebrón2005}%
  \BibitemOpen
  \bibfield  {author} {\bibinfo {author} {\bibfnamefont {J.~A.}\ \bibnamefont {Acebrón}}, \bibinfo {author} {\bibfnamefont {L.~L.}\ \bibnamefont {Bonilla}}, \bibinfo {author} {\bibfnamefont {C.~J.}\ \bibnamefont {Pérez~Vicente}}, \bibinfo {author} {\bibfnamefont {F.}~\bibnamefont {Ritort}},\ and\ \bibinfo {author} {\bibfnamefont {R.}~\bibnamefont {Spigler}},\ }\bibfield  {title} {\bibinfo {title} {The kuramoto model: A simple paradigm for synchronization phenomena},\ }\href {https://doi.org/10.1103/revmodphys.77.137} {\bibfield  {journal} {\bibinfo  {journal} {Rev. Mod. Phys.}\ }\textbf {\bibinfo {volume} {77}},\ \bibinfo {pages} {137–185} (\bibinfo {year} {2005})}\BibitemShut {NoStop}%
\bibitem [{\citenamefont {Rosenblum}\ and\ \citenamefont {Pikovsky}(2004)}]{Rosenblum2004}%
  \BibitemOpen
  \bibfield  {author} {\bibinfo {author} {\bibfnamefont {M.~G.}\ \bibnamefont {Rosenblum}}\ and\ \bibinfo {author} {\bibfnamefont {A.~S.}\ \bibnamefont {Pikovsky}},\ }\bibfield  {title} {\bibinfo {title} {Controlling synchronization in an ensemble of globally coupled oscillators},\ }\bibfield  {journal} {\bibinfo  {journal} {Phys. Rev. Lett.}\ }\textbf {\bibinfo {volume} {92}},\ \href {https://doi.org/10.1103/physrevlett.92.114102} {10.1103/physrevlett.92.114102} (\bibinfo {year} {2004})\BibitemShut {NoStop}%
\bibitem [{\citenamefont {Tukhlina}\ \emph {et~al.}(2008)\citenamefont {Tukhlina}, \citenamefont {Rosenblum},\ and\ \citenamefont {Pikovsky}}]{Tukhlina2008}%
  \BibitemOpen
  \bibfield  {author} {\bibinfo {author} {\bibfnamefont {N.}~\bibnamefont {Tukhlina}}, \bibinfo {author} {\bibfnamefont {M.}~\bibnamefont {Rosenblum}},\ and\ \bibinfo {author} {\bibfnamefont {A.}~\bibnamefont {Pikovsky}},\ }\bibfield  {title} {\bibinfo {title} {Controlling coherence of noisy and chaotic oscillators by a linear feedback},\ }\href {https://doi.org/10.1016/j.physa.2008.06.054} {\bibfield  {journal} {\bibinfo  {journal} {Phys. A}\ }\textbf {\bibinfo {volume} {387}},\ \bibinfo {pages} {6045–6056} (\bibinfo {year} {2008})}\BibitemShut {NoStop}%
\bibitem [{\citenamefont {Goldobin}\ \emph {et~al.}(2003)\citenamefont {Goldobin}, \citenamefont {Rosenblum},\ and\ \citenamefont {Pikovsky}}]{Goldobin2003}%
  \BibitemOpen
  \bibfield  {author} {\bibinfo {author} {\bibfnamefont {D.}~\bibnamefont {Goldobin}}, \bibinfo {author} {\bibfnamefont {M.}~\bibnamefont {Rosenblum}},\ and\ \bibinfo {author} {\bibfnamefont {A.}~\bibnamefont {Pikovsky}},\ }\bibfield  {title} {\bibinfo {title} {Coherence of noisy oscillators with delayed feedback},\ }\href {https://doi.org/10.1016/s0378-4371(03)00463-1} {\bibfield  {journal} {\bibinfo  {journal} {Phys. A}\ }\textbf {\bibinfo {volume} {327}},\ \bibinfo {pages} {124–128} (\bibinfo {year} {2003})}\BibitemShut {NoStop}%
\bibitem [{\citenamefont {Aranson}\ and\ \citenamefont {Kramer}(2002)}]{Aranson2002}%
  \BibitemOpen
  \bibfield  {author} {\bibinfo {author} {\bibfnamefont {I.~S.}\ \bibnamefont {Aranson}}\ and\ \bibinfo {author} {\bibfnamefont {L.}~\bibnamefont {Kramer}},\ }\bibfield  {title} {\bibinfo {title} {The world of the complex ginzburg-landau equation},\ }\href {https://doi.org/10.1103/revmodphys.74.99} {\bibfield  {journal} {\bibinfo  {journal} {Rev. Mod. Phys.}\ }\textbf {\bibinfo {volume} {74}},\ \bibinfo {pages} {99–143} (\bibinfo {year} {2002})}\BibitemShut {NoStop}%
\bibitem [{\citenamefont {Hohenberg}\ and\ \citenamefont {Halperin}(1977)}]{Hohenberg1977}%
  \BibitemOpen
  \bibfield  {author} {\bibinfo {author} {\bibfnamefont {P.~C.}\ \bibnamefont {Hohenberg}}\ and\ \bibinfo {author} {\bibfnamefont {B.~I.}\ \bibnamefont {Halperin}},\ }\bibfield  {title} {\bibinfo {title} {Theory of dynamic critical phenomena},\ }\href {https://doi.org/10.1103/revmodphys.49.435} {\bibfield  {journal} {\bibinfo  {journal} {Rev. Mod. Phys.}\ }\textbf {\bibinfo {volume} {49}},\ \bibinfo {pages} {435–479} (\bibinfo {year} {1977})}\BibitemShut {NoStop}%
\bibitem [{\citenamefont {Surowka}\ and\ \citenamefont {Witkowski}(2018)}]{Surowka2018}%
  \BibitemOpen
  \bibfield  {author} {\bibinfo {author} {\bibfnamefont {P.}~\bibnamefont {Surowka}}\ and\ \bibinfo {author} {\bibfnamefont {P.}~\bibnamefont {Witkowski}},\ }\bibfield  {title} {\bibinfo {title} {Symmetries in the path integral formulation of the langevin dynamics},\ }\bibfield  {journal} {\bibinfo  {journal} {Phys. Rev. E}\ }\textbf {\bibinfo {volume} {98}},\ \href {https://doi.org/10.1103/physreve.98.042140} {10.1103/physreve.98.042140} (\bibinfo {year} {2018})\BibitemShut {NoStop}%
\bibitem [{\citenamefont {Sieberer}\ \emph {et~al.}(2015)\citenamefont {Sieberer}, \citenamefont {Chiocchetta}, \citenamefont {Gambassi}, \citenamefont {Täuber},\ and\ \citenamefont {Diehl}}]{Sieberer2015}%
  \BibitemOpen
  \bibfield  {author} {\bibinfo {author} {\bibfnamefont {L.~M.}\ \bibnamefont {Sieberer}}, \bibinfo {author} {\bibfnamefont {A.}~\bibnamefont {Chiocchetta}}, \bibinfo {author} {\bibfnamefont {A.}~\bibnamefont {Gambassi}}, \bibinfo {author} {\bibfnamefont {U.~C.}\ \bibnamefont {Täuber}},\ and\ \bibinfo {author} {\bibfnamefont {S.}~\bibnamefont {Diehl}},\ }\bibfield  {title} {\bibinfo {title} {Thermodynamic equilibrium as a symmetry of the schwinger-keldysh action},\ }\bibfield  {journal} {\bibinfo  {journal} {Phys. Rev. B}\ }\textbf {\bibinfo {volume} {92}},\ \href {https://doi.org/10.1103/physrevb.92.134307} {10.1103/physrevb.92.134307} (\bibinfo {year} {2015})\BibitemShut {NoStop}%
\bibitem [{\citenamefont {Kardar}\ \emph {et~al.}(1986)\citenamefont {Kardar}, \citenamefont {Parisi},\ and\ \citenamefont {Zhang}}]{Kardar1986}%
  \BibitemOpen
  \bibfield  {author} {\bibinfo {author} {\bibfnamefont {M.}~\bibnamefont {Kardar}}, \bibinfo {author} {\bibfnamefont {G.}~\bibnamefont {Parisi}},\ and\ \bibinfo {author} {\bibfnamefont {Y.-C.}\ \bibnamefont {Zhang}},\ }\bibfield  {title} {\bibinfo {title} {Dynamic scaling of growing interfaces},\ }\href {https://doi.org/10.1103/physrevlett.56.889} {\bibfield  {journal} {\bibinfo  {journal} {Phys. Rev. Lett.}\ }\textbf {\bibinfo {volume} {56}},\ \bibinfo {pages} {889–892} (\bibinfo {year} {1986})}\BibitemShut {NoStop}%
\bibitem [{\citenamefont {Weis}\ \emph {et~al.}(2023)\citenamefont {Weis}, \citenamefont {Fruchart}, \citenamefont {Hanai}, \citenamefont {Kawagoe}, \citenamefont {Littlewood},\ and\ \citenamefont {Vitelli}}]{Weis2023}%
  \BibitemOpen
  \bibfield  {author} {\bibinfo {author} {\bibfnamefont {C.}~\bibnamefont {Weis}}, \bibinfo {author} {\bibfnamefont {M.}~\bibnamefont {Fruchart}}, \bibinfo {author} {\bibfnamefont {R.}~\bibnamefont {Hanai}}, \bibinfo {author} {\bibfnamefont {K.}~\bibnamefont {Kawagoe}}, \bibinfo {author} {\bibfnamefont {P.~B.}\ \bibnamefont {Littlewood}},\ and\ \bibinfo {author} {\bibfnamefont {V.}~\bibnamefont {Vitelli}},\ }\bibfield  {title} {\bibinfo {title} {Exceptional points in nonlinear and stochastic dynamics},\ }\href@noop {} {\  (\bibinfo {year} {2023})},\ \Eprint {https://arxiv.org/abs/2207.11667} {arXiv:2207.11667 [nlin.CD]} \BibitemShut {NoStop}%
\bibitem [{\citenamefont {Fruchart}\ \emph {et~al.}(2021)\citenamefont {Fruchart}, \citenamefont {Hanai}, \citenamefont {Littlewood},\ and\ \citenamefont {Vitelli}}]{Fruchart2021}%
  \BibitemOpen
  \bibfield  {author} {\bibinfo {author} {\bibfnamefont {M.}~\bibnamefont {Fruchart}}, \bibinfo {author} {\bibfnamefont {R.}~\bibnamefont {Hanai}}, \bibinfo {author} {\bibfnamefont {P.~B.}\ \bibnamefont {Littlewood}},\ and\ \bibinfo {author} {\bibfnamefont {V.}~\bibnamefont {Vitelli}},\ }\bibfield  {title} {\bibinfo {title} {Non-reciprocal phase transitions},\ }\href {https://doi.org/10.1038/s41586-021-03375-9} {\bibfield  {journal} {\bibinfo  {journal} {Nature}\ }\textbf {\bibinfo {volume} {592}},\ \bibinfo {pages} {363–369} (\bibinfo {year} {2021})}\BibitemShut {NoStop}%
\bibitem [{\citenamefont {Hanai}\ and\ \citenamefont {Littlewood}(2020)}]{Hanai2020}%
  \BibitemOpen
  \bibfield  {author} {\bibinfo {author} {\bibfnamefont {R.}~\bibnamefont {Hanai}}\ and\ \bibinfo {author} {\bibfnamefont {P.~B.}\ \bibnamefont {Littlewood}},\ }\bibfield  {title} {\bibinfo {title} {Critical fluctuations at a many-body exceptional point},\ }\bibfield  {journal} {\bibinfo  {journal} {Phys. Rev. Res.}\ }\textbf {\bibinfo {volume} {2}},\ \href {https://doi.org/10.1103/physrevresearch.2.033018} {10.1103/physrevresearch.2.033018} (\bibinfo {year} {2020})\BibitemShut {NoStop}%
\bibitem [{\citenamefont {Hanai}\ \emph {et~al.}(2019)\citenamefont {Hanai}, \citenamefont {Edelman}, \citenamefont {Ohashi},\ and\ \citenamefont {Littlewood}}]{Hanai2019}%
  \BibitemOpen
  \bibfield  {author} {\bibinfo {author} {\bibfnamefont {R.}~\bibnamefont {Hanai}}, \bibinfo {author} {\bibfnamefont {A.}~\bibnamefont {Edelman}}, \bibinfo {author} {\bibfnamefont {Y.}~\bibnamefont {Ohashi}},\ and\ \bibinfo {author} {\bibfnamefont {P.~B.}\ \bibnamefont {Littlewood}},\ }\bibfield  {title} {\bibinfo {title} {Non-hermitian phase transition from a polariton bose-einstein condensate to a photon laser},\ }\bibfield  {journal} {\bibinfo  {journal} {Phys. Rev. Lett.}\ }\textbf {\bibinfo {volume} {122}},\ \href {https://doi.org/10.1103/physrevlett.122.185301} {10.1103/physrevlett.122.185301} (\bibinfo {year} {2019})\BibitemShut {NoStop}%
\bibitem [{\citenamefont {Chen}\ \emph {et~al.}(2017{\natexlab{a}})\citenamefont {Chen}, \citenamefont {Kaya~Ozdemir}, \citenamefont {Zhao}, \citenamefont {Wiersig},\ and\ \citenamefont {Yang}}]{Chen2017}%
  \BibitemOpen
  \bibfield  {author} {\bibinfo {author} {\bibfnamefont {W.}~\bibnamefont {Chen}}, \bibinfo {author} {\bibfnamefont {S.}~\bibnamefont {Kaya~Ozdemir}}, \bibinfo {author} {\bibfnamefont {G.}~\bibnamefont {Zhao}}, \bibinfo {author} {\bibfnamefont {J.}~\bibnamefont {Wiersig}},\ and\ \bibinfo {author} {\bibfnamefont {L.}~\bibnamefont {Yang}},\ }\bibfield  {title} {\bibinfo {title} {Exceptional points enhance sensing in an optical microcavity},\ }\href {https://doi.org/10.1038/nature23281} {\bibfield  {journal} {\bibinfo  {journal} {Nature}\ }\textbf {\bibinfo {volume} {548}},\ \bibinfo {pages} {192–196} (\bibinfo {year} {2017}{\natexlab{a}})}\BibitemShut {NoStop}%
\bibitem [{\citenamefont {Zhang}\ \emph {et~al.}(2019{\natexlab{a}})\citenamefont {Zhang}, \citenamefont {Sweeney}, \citenamefont {Hsu}, \citenamefont {Yang}, \citenamefont {Stone},\ and\ \citenamefont {Jiang}}]{Zhang2019}%
  \BibitemOpen
  \bibfield  {author} {\bibinfo {author} {\bibfnamefont {M.}~\bibnamefont {Zhang}}, \bibinfo {author} {\bibfnamefont {W.}~\bibnamefont {Sweeney}}, \bibinfo {author} {\bibfnamefont {C.~W.}\ \bibnamefont {Hsu}}, \bibinfo {author} {\bibfnamefont {L.}~\bibnamefont {Yang}}, \bibinfo {author} {\bibfnamefont {A.}~\bibnamefont {Stone}},\ and\ \bibinfo {author} {\bibfnamefont {L.}~\bibnamefont {Jiang}},\ }\bibfield  {title} {\bibinfo {title} {Quantum noise theory of exceptional point amplifying sensors},\ }\bibfield  {journal} {\bibinfo  {journal} {Phys. Rev. Lett.}\ }\textbf {\bibinfo {volume} {123}},\ \href {https://doi.org/10.1103/physrevlett.123.180501} {10.1103/physrevlett.123.180501} (\bibinfo {year} {2019}{\natexlab{a}})\BibitemShut {NoStop}%
\bibitem [{\citenamefont {Shmakov}\ and\ \citenamefont {Littlewood}(2024)}]{Shmakov2024}%
  \BibitemOpen
  \bibfield  {author} {\bibinfo {author} {\bibfnamefont {S.}~\bibnamefont {Shmakov}}\ and\ \bibinfo {author} {\bibfnamefont {P.~B.}\ \bibnamefont {Littlewood}},\ }\bibfield  {title} {\bibinfo {title} {Coalescence of limit cycles in the presence of noise},\ }\bibfield  {journal} {\bibinfo  {journal} {Phys. Rev. E}\ }\textbf {\bibinfo {volume} {109}},\ \href {https://doi.org/10.1103/physreve.109.024220} {10.1103/physreve.109.024220} (\bibinfo {year} {2024})\BibitemShut {NoStop}%
\bibitem [{\citenamefont {Shmakov}\ \emph {et~al.}(2024)\citenamefont {Shmakov}, \citenamefont {Osipycheva},\ and\ \citenamefont {Littlewood}}]{Shmakov2024b}%
  \BibitemOpen
  \bibfield  {author} {\bibinfo {author} {\bibfnamefont {S.}~\bibnamefont {Shmakov}}, \bibinfo {author} {\bibfnamefont {G.}~\bibnamefont {Osipycheva}},\ and\ \bibinfo {author} {\bibfnamefont {P.~B.}\ \bibnamefont {Littlewood}},\ }\bibfield  {title} {\bibinfo {title} {Gaussian fluctuations of non-reciprocal systems},\ }\href {https://arxiv.org/abs/2411.17944} {\  (\bibinfo {year} {2024})},\ \Eprint {https://arxiv.org/abs/2411.17944} {arXiv:2411.17944 [cond-mat.stat-mech]} \BibitemShut {NoStop}%
\bibitem [{\citenamefont {Henry}(1982)}]{Henry1982}%
  \BibitemOpen
  \bibfield  {author} {\bibinfo {author} {\bibfnamefont {C.}~\bibnamefont {Henry}},\ }\bibfield  {title} {\bibinfo {title} {Theory of the linewidth of semiconductor lasers},\ }\href {https://doi.org/10.1109/jqe.1982.1071522} {\bibfield  {journal} {\bibinfo  {journal} {IEEE J. Quantum Electron.}\ }\textbf {\bibinfo {volume} {18}},\ \bibinfo {pages} {259–264} (\bibinfo {year} {1982})}\BibitemShut {NoStop}%
\bibitem [{\citenamefont {Lauter}\ \emph {et~al.}(2015)\citenamefont {Lauter}, \citenamefont {Brendel}, \citenamefont {Habraken},\ and\ \citenamefont {Marquardt}}]{Lauter2015}%
  \BibitemOpen
  \bibfield  {author} {\bibinfo {author} {\bibfnamefont {R.}~\bibnamefont {Lauter}}, \bibinfo {author} {\bibfnamefont {C.}~\bibnamefont {Brendel}}, \bibinfo {author} {\bibfnamefont {S.~J.}\ \bibnamefont {Habraken}},\ and\ \bibinfo {author} {\bibfnamefont {F.}~\bibnamefont {Marquardt}},\ }\bibfield  {title} {\bibinfo {title} {Pattern phase diagram for two-dimensional arrays of coupled limit-cycle oscillators},\ }\bibfield  {journal} {\bibinfo  {journal} {Phys. Rev. E}\ }\textbf {\bibinfo {volume} {92}},\ \href {https://doi.org/10.1103/physreve.92.012902} {10.1103/physreve.92.012902} (\bibinfo {year} {2015})\BibitemShut {NoStop}%
\bibitem [{\citenamefont {Canolty}\ and\ \citenamefont {Knight}(2010)}]{Canolty2010}%
  \BibitemOpen
  \bibfield  {author} {\bibinfo {author} {\bibfnamefont {R.~T.}\ \bibnamefont {Canolty}}\ and\ \bibinfo {author} {\bibfnamefont {R.~T.}\ \bibnamefont {Knight}},\ }\bibfield  {title} {\bibinfo {title} {The functional role of cross-frequency coupling},\ }\href {https://doi.org/10.1016/j.tics.2010.09.001} {\bibfield  {journal} {\bibinfo  {journal} {Trends Cognit. Sci.}\ }\textbf {\bibinfo {volume} {14}},\ \bibinfo {pages} {506–515} (\bibinfo {year} {2010})}\BibitemShut {NoStop}%
\bibitem [{\citenamefont {Hertz}\ \emph {et~al.}(2016)\citenamefont {Hertz}, \citenamefont {Roudi},\ and\ \citenamefont {Sollich}}]{Hertz2016}%
  \BibitemOpen
  \bibfield  {author} {\bibinfo {author} {\bibfnamefont {J.~A.}\ \bibnamefont {Hertz}}, \bibinfo {author} {\bibfnamefont {Y.}~\bibnamefont {Roudi}},\ and\ \bibinfo {author} {\bibfnamefont {P.}~\bibnamefont {Sollich}},\ }\bibfield  {title} {\bibinfo {title} {Path integral methods for the dynamics of stochastic and disordered systems},\ }\href {https://doi.org/10.1088/1751-8121/50/3/033001} {\bibfield  {journal} {\bibinfo  {journal} {J. Phys. A: Math. Theor.}\ }\textbf {\bibinfo {volume} {50}},\ \bibinfo {pages} {033001} (\bibinfo {year} {2016})}\BibitemShut {NoStop}%
\bibitem [{\citenamefont {Gardiner}(1997)}]{Gardiner1997}%
  \BibitemOpen
  \bibfield  {author} {\bibinfo {author} {\bibfnamefont {C.~W.}\ \bibnamefont {Gardiner}},\ }\href@noop {} {\emph {\bibinfo {title} {Handbook of Stochastic Methods: For Physics, chemistry and natural sciences}}}\ (\bibinfo  {publisher} {Springer},\ \bibinfo {year} {1997})\BibitemShut {NoStop}%
\bibitem [{\citenamefont {Martin}\ \emph {et~al.}(1973)\citenamefont {Martin}, \citenamefont {Siggia},\ and\ \citenamefont {Rose}}]{Martin1973}%
  \BibitemOpen
  \bibfield  {author} {\bibinfo {author} {\bibfnamefont {P.~C.}\ \bibnamefont {Martin}}, \bibinfo {author} {\bibfnamefont {E.~D.}\ \bibnamefont {Siggia}},\ and\ \bibinfo {author} {\bibfnamefont {H.~A.}\ \bibnamefont {Rose}},\ }\bibfield  {title} {\bibinfo {title} {Statistical dynamics of classical systems},\ }\href {https://doi.org/10.1103/physreva.8.423} {\bibfield  {journal} {\bibinfo  {journal} {Phys. Rev. A}\ }\textbf {\bibinfo {volume} {8}},\ \bibinfo {pages} {423–437} (\bibinfo {year} {1973})}\BibitemShut {NoStop}%
\bibitem [{\citenamefont {De~Dominicis}\ and\ \citenamefont {Peliti}(1978)}]{DeDominicis1978}%
  \BibitemOpen
  \bibfield  {author} {\bibinfo {author} {\bibfnamefont {C.}~\bibnamefont {De~Dominicis}}\ and\ \bibinfo {author} {\bibfnamefont {L.}~\bibnamefont {Peliti}},\ }\bibfield  {title} {\bibinfo {title} {Field-theory renormalization and critical dynamics above $\mathrm{T}_c$: Helium, antiferro- magnets, and liquid-gas systems},\ }\href {https://doi.org/10.1103/physrevb.18.353} {\bibfield  {journal} {\bibinfo  {journal} {Phys. Rev. B}\ }\textbf {\bibinfo {volume} {18}},\ \bibinfo {pages} {353–376} (\bibinfo {year} {1978})}\BibitemShut {NoStop}%
\bibitem [{\citenamefont {Biancalani}\ \emph {et~al.}(2017)\citenamefont {Biancalani}, \citenamefont {Jafarpour},\ and\ \citenamefont {Goldenfeld}}]{Biancalani2017}%
  \BibitemOpen
  \bibfield  {author} {\bibinfo {author} {\bibfnamefont {T.}~\bibnamefont {Biancalani}}, \bibinfo {author} {\bibfnamefont {F.}~\bibnamefont {Jafarpour}},\ and\ \bibinfo {author} {\bibfnamefont {N.}~\bibnamefont {Goldenfeld}},\ }\bibfield  {title} {\bibinfo {title} {Giant amplification of noise in fluctuation-induced pattern formation},\ }\bibfield  {journal} {\bibinfo  {journal} {Phys. Rev. Lett.}\ }\textbf {\bibinfo {volume} {118}},\ \href {https://doi.org/10.1103/physrevlett.118.018101} {10.1103/physrevlett.118.018101} (\bibinfo {year} {2017})\BibitemShut {NoStop}%
\bibitem [{\citenamefont {Chen}\ \emph {et~al.}(2017{\natexlab{b}})\citenamefont {Chen}, \citenamefont {Kaya~Ozdemir}, \citenamefont {Zhao}, \citenamefont {Wiersig},\ and\ \citenamefont {Yang}}]{Chen2017_2}%
  \BibitemOpen
  \bibfield  {author} {\bibinfo {author} {\bibfnamefont {W.}~\bibnamefont {Chen}}, \bibinfo {author} {\bibfnamefont {S.}~\bibnamefont {Kaya~Ozdemir}}, \bibinfo {author} {\bibfnamefont {G.}~\bibnamefont {Zhao}}, \bibinfo {author} {\bibfnamefont {J.}~\bibnamefont {Wiersig}},\ and\ \bibinfo {author} {\bibfnamefont {L.}~\bibnamefont {Yang}},\ }\bibfield  {title} {\bibinfo {title} {Exceptional points enhance sensing in an optical microcavity},\ }\href {https://doi.org/10.1038/nature23281} {\bibfield  {journal} {\bibinfo  {journal} {Nature}\ }\textbf {\bibinfo {volume} {548}},\ \bibinfo {pages} {192–196} (\bibinfo {year} {2017}{\natexlab{b}})}\BibitemShut {NoStop}%
\bibitem [{\citenamefont {Zhang}\ \emph {et~al.}(2019{\natexlab{b}})\citenamefont {Zhang}, \citenamefont {Sweeney}, \citenamefont {Hsu}, \citenamefont {Yang}, \citenamefont {Stone},\ and\ \citenamefont {Jiang}}]{Zhang2019_2}%
  \BibitemOpen
  \bibfield  {author} {\bibinfo {author} {\bibfnamefont {M.}~\bibnamefont {Zhang}}, \bibinfo {author} {\bibfnamefont {W.}~\bibnamefont {Sweeney}}, \bibinfo {author} {\bibfnamefont {C.~W.}\ \bibnamefont {Hsu}}, \bibinfo {author} {\bibfnamefont {L.}~\bibnamefont {Yang}}, \bibinfo {author} {\bibfnamefont {A.}~\bibnamefont {Stone}},\ and\ \bibinfo {author} {\bibfnamefont {L.}~\bibnamefont {Jiang}},\ }\bibfield  {title} {\bibinfo {title} {Quantum noise theory of exceptional point amplifying sensors},\ }\bibfield  {journal} {\bibinfo  {journal} {Phys. Rev. Lett.}\ }\textbf {\bibinfo {volume} {123}},\ \href {https://doi.org/10.1103/physrevlett.123.180501} {10.1103/physrevlett.123.180501} (\bibinfo {year} {2019}{\natexlab{b}})\BibitemShut {NoStop}%
\bibitem [{\citenamefont {Huang}\ \emph {et~al.}(2025)\citenamefont {Huang}, \citenamefont {Hanai},\ and\ \citenamefont {Littlewood}}]{huang25}%
  \BibitemOpen
  \bibfield  {author} {\bibinfo {author} {\bibfnamefont {X.}~\bibnamefont {Huang}}, \bibinfo {author} {\bibfnamefont {R.}~\bibnamefont {Hanai}},\ and\ \bibinfo {author} {\bibfnamefont {P.}~\bibnamefont {Littlewood}},\ }\bibfield  {title} {\bibinfo {title} {Nonlinear dynamics in a polariton condensate},\ }\href@noop {} {\bibfield  {journal} {\bibinfo  {journal} {unpublished}\ } (\bibinfo {year} {2025})}\BibitemShut {NoStop}%
\bibitem [{\citenamefont {Mermin}\ and\ \citenamefont {Wagner}(1966)}]{Mermin1966}%
  \BibitemOpen
  \bibfield  {author} {\bibinfo {author} {\bibfnamefont {N.~D.}\ \bibnamefont {Mermin}}\ and\ \bibinfo {author} {\bibfnamefont {H.}~\bibnamefont {Wagner}},\ }\bibfield  {title} {\bibinfo {title} {Absence of ferromagnetism or antiferromagnetism in one- or two-dimensional isotropic heisenberg models},\ }\href {https://doi.org/10.1103/physrevlett.17.1133} {\bibfield  {journal} {\bibinfo  {journal} {Phys. Rev. Lett.}\ }\textbf {\bibinfo {volume} {17}},\ \bibinfo {pages} {1133–1136} (\bibinfo {year} {1966})}\BibitemShut {NoStop}%
\bibitem [{\citenamefont {Daviet}\ \emph {et~al.}(2024{\natexlab{b}})\citenamefont {Daviet}, \citenamefont {Zelle}, \citenamefont {Asadollahi},\ and\ \citenamefont {Diehl}}]{Daviet2024b}%
  \BibitemOpen
  \bibfield  {author} {\bibinfo {author} {\bibfnamefont {R.}~\bibnamefont {Daviet}}, \bibinfo {author} {\bibfnamefont {C.~P.}\ \bibnamefont {Zelle}}, \bibinfo {author} {\bibfnamefont {A.}~\bibnamefont {Asadollahi}},\ and\ \bibinfo {author} {\bibfnamefont {S.}~\bibnamefont {Diehl}},\ }\bibfield  {title} {\bibinfo {title} {Kardar-parisi-zhang scaling in time-crystalline matter},\ }\href {https://arxiv.org/abs/2412.09677} {\  (\bibinfo {year} {2024}{\natexlab{b}})},\ \Eprint {https://arxiv.org/abs/2412.09677} {arXiv:2412.09677 [cond-mat.stat-mech]} \BibitemShut {NoStop}%
\bibitem [{\citenamefont {Sieberer}\ \emph {et~al.}(2016)\citenamefont {Sieberer}, \citenamefont {Wachtel}, \citenamefont {Altman},\ and\ \citenamefont {Diehl}}]{Sieberer2016}%
  \BibitemOpen
  \bibfield  {author} {\bibinfo {author} {\bibfnamefont {L.~M.}\ \bibnamefont {Sieberer}}, \bibinfo {author} {\bibfnamefont {G.}~\bibnamefont {Wachtel}}, \bibinfo {author} {\bibfnamefont {E.}~\bibnamefont {Altman}},\ and\ \bibinfo {author} {\bibfnamefont {S.}~\bibnamefont {Diehl}},\ }\bibfield  {title} {\bibinfo {title} {Lattice duality for the compact kardar-parisi-zhang equation},\ }\bibfield  {journal} {\bibinfo  {journal} {Phys. Rev. B}\ }\textbf {\bibinfo {volume} {94}},\ \href {https://doi.org/10.1103/physrevb.94.104521} {10.1103/physrevb.94.104521} (\bibinfo {year} {2016})\BibitemShut {NoStop}%
\end{thebibliography}%

\end{document}